# Metacognitive Skill Learning:
# A Computational Account

by

Brendan Conway-Smith

A thesis submitted to the Office of Graduate Studies
in partial fulfillment of the requirements for the degree of

Doctor of Philosophy

in

Cognitive Science

Carleton University
Ottawa, Ontario

## Abstract

This dissertation presents the first formal theory of metacognitive skill learning. Metacognition, the capacity to monitor and control one's own cognitive processes, has been widely studied, yet the field still lacks a theoretical framework explaining how metacognitive abilities are learned. This gap limits progress in both theory and application across fields such as cognitive science, education, therapeutic practice, and artificial intelligence.

The account developed here builds on classic models of skill acquisition from perceptual-motor and cognitive domains. It proposes that metacognitive skill develops primarily through proceduralization, whereby explicit, effortful metacognitive processes are transformed through practice into implicit, automatic routines.

The explanatory power of this theory is shown by its ability to unify diverse phenomena, including attentional training, emotion regulation, the metacognitive threshold, and detached mindfulness. The result is an integrated, mechanistic account of metacognitive skill that organizes existing findings, generates testable predictions, informs the design of AI, and supports the strengthening of metacognitive skill in everyday life.

## Acknowledgements

I am grateful to Dr. Robert West for his supervision and support throughout this project, and to Dr. Myrto Mylopoulos for her encouragement and insightful feedback, which have profoundly shaped the ideas in this thesis. To the faculty and staff in the Department of Cognitive Science, thank you for giving me a place to stand.

# Table of Contents

## List of Tables

## List of Figures

***Table 1.*** *Glossary of key metacognitive terms*

| **Glossary** | |
|---|---|
| **Term** | **Definition** |
| **Metacognition** | The capacity of cognition to monitor and control its own processes, where mental states and operations become objects of awareness, representation, evaluation, and regulation. |
| **Metacognitive monitoring** | The capacity to represent the properties of cognitive states and processes (e.g., thoughts, emotions, memory, feelings) for the purpose of control. |
| **Metacognitive control** | The regulation of cognitive processes through mental actions (e.g., directing attention, regulating emotion, strategic learning) in the service of internally-directed, explicit metacognitive goals. |
| **Cognitive control** | The regulation of cognitive processes (e.g., attention, memory, reasoning) instrumentally in the service of externally-directed goals. |
| **Metacognitive vs. Cognitive control** | A distinction in which cognitive control regulates cognition instrumentally in service of external goals, whereas metacognitive control regulates cognition toward metacognitive goals as ends in themselves. |
| **Metacognitive skill** | The learned capacity to reliably and flexibly monitor and regulate one's cognitive activity across contexts to achieve metacognitive goals, enabling effective control over attention, emotion, memory, reasoning, and thought. |
| **Object-level vs. Meta-level** | Object-level cognitive processes operate on the external world. Higher-order metacognitive processes monitor and regulate cognitive processes. |
| **Declarative metacognitive knowledge** | Propositional knowledge about cognitive processes (metarepresentations), including explicit models, strategies, and facts about how cognition operates and which mental actions achieve metacognitive goals. |
| **Procedural metacognitive knowledge** | Implicit, non-verbal knowledge that enables the execution of metacognitive actions efficiently and automatically, often without conscious access. |
| **Metacognitive feelings** | Non-conceptual, analog, affective signals (e.g., confidence, feeling of knowing) that guide cognition implicitly without propositional format. |
| **Metacognitive goals** | Proximal, propositional goals whose content depicts a mental process or state (e.g., attention, emotion, memory, learning, reasoning) as an end in itself. |
| **Metacognitive proceduralization /automatization** | The process by which metacognitive control becomes faster, more effective, fluid, and automatic. |
| **Dual-process metacognition** | A distinction between Type 2 metacognition (explicit, slow, conceptually-driven) and Type 1 metacognition (implicit, fast, automatic), with skill learning partly involving a transition from the former to the latter. |
| **Type 1 metacognition** (procedural/ model-free) | A form of metacognition that operates through fast, automatic, and non-conceptual processes, typically driven by affective or heuristic signals (e.g., feelings of knowing, confidence). It monitors and guides cognition implicitly, without relying on explicit representations of one's own mental states, and is often not directly accessible to conscious report. |
| **Type 2 metacognition** (analytic/ model-based) | A form of deliberate metacognition that involves slow, effortful, and explicitly represented knowledge about one's own cognitive processes. It relies on metarepresentations with conceptual content, such as beliefs about one's attention or reasoning, or instructions that guide monitoring and control processes toward metacognitive goals. |
| **Production rules** | Procedural knowledge as computationally formatted. Condition–action pairings ("if–then" rules) that execute metacognitive control. |
| **Gradability** | A three-dimensional account of skill variation: success rate (height), range of achievable goals (breadth), and adaptability across contexts (depth). |
| **Metacognitive flexibility** | The capacity to adapt monitoring and control processes across changing internal states and environments, distinguishing skill from rigid habit. |
| **Metacognitive threshold** | The minimum level of a mental stimulus required for it to become available to awareness and subject to metacognitive monitoring and control. Lowering the threshold through training increases metacognitive sensitivity, enabling finer perception and discernment of internal cognitive states. |
| **Detached mindfulness** | A skill-based metacognitive stance involving non-reactive monitoring of thoughts, emotions, and bodily sensations, enabling adaptive disengagement from maladaptive cognitive patterns. It develops through attentional refinement and increased metacognitive sensitivity to subtle, momentary changes in affective experience. |

***Table 2.*** *Contributions of the dissertation*

| Contributions | Type | Location |
|---|---|---|
| Characterizes metacognition as a domain of skill with identifiable goals, structures, and constraints. | Theoretical | Chapter 2 |
| Proposes a proceduralization-based account of metacognitive skill acquisition. | Theoretical / Computational | Chapter 3 |
| Provides an account of metacognitive monitoring and control using a computational cognitive architecture. | Computational | Chapters 3–6 |
| Applies the proceduralization framework developed in Chapter 3 to explain attentional training, emotion regulation, and other metacognitive phenomena. | Theoretical / Applied | Chapters 4–6 |
| Introduces and models concepts such as the metacognitive threshold and detached mindfulness. | Theoretical / Computational | Chapters 5–6 |
| Derives testable predictions about metacognitive skill learning (e.g., proceduralization effects). | Theoretical / Empirical | Chapter 7 |
| Develops working computational models of metacognitive processes (GitHub repository). | Methodological | Repository |

## Contributions of Authors

The author of this dissertation, Brendan Conway-Smith, conceived the central ideas presented herein and carried out the work under the supervision of Dr. Robert L. West. *Chapter 2 (Manuscript 1):* Sole-authored by Brendan Conway-Smith. *Chapter 3 (Manuscript 2):* Brendan Conway-Smith conceived the ideas and wrote the manuscript. Dr. Robert L. West and Dr. Myrto Mylopoulos provided consultation, reviewed and edited the manuscript, and provided feedback. *Chapters 4–6 (Manuscripts 3–5):* Brendan Conway-Smith conceived the ideas and wrote the manuscripts. Dr. Robert L. West reviewed the manuscripts and provided feedback.

## Declaration

This dissertation is my original work. All sources have been appropriately acknowledged. Any material included in manuscript form has been identified, along with the contributions of co-authors and supervisors. Unless otherwise stated, the work presented in this dissertation was conceived and carried out by me.

## Chapter Overview

The organization of this manuscript-based thesis adheres to the thesis preparation guidelines set out by Carleton University Graduate and Postdoctoral Studies.

**Chapter 1 - Introduction**

Outlines the central claims, theoretical methodology, and overall framework. Manuscript chapters contain their own literature reviews.

**Chapter 2 (Manuscript) - Metacognition as a Domain of Skill**

Provides the conceptual groundwork for treating metacognition as a domain of skill. Drawing on philosophy of action and skill acquisition research, it identifies key features of skilled action and examines how they apply to metacognition. This framework clarifies metacognitive skill as the phenomenon that later chapters aim to explain computationally.

**Chapter 3 (Manuscript) - Metacognitive Skill: How It Is Acquired**

*Presents the core theoretical proposal,* arguing that metacognitive skill learning is largely explained by the Fitts-Anderson model of proceduralization. It introduces a framework of metacognitive proceduralization that clarifies the cognitive mechanisms underlying metacognitive skill learning and helps account for empirical findings.

**Chapter 4 (Manuscript) - Metacognitive Mechanisms of the Attentional Training Technique**

Applies the metacognitive proceduralization framework proposed in Chapter 3 to the Attentional Training Technique (ATT), arguing that ATT transforms declarative attentional strategies into automatic procedural skills, thereby enhancing metacognitive control and emotion regulation.

**Chapter 5 (Manuscript) - Metacognitive Threshold: A Computational Account**

Applies the same skill framework proposed in Chapter 3 to model the refinement of the metacognitive threshold, understood as the minimum stimulus needed for a mental state to be perceived, and explains how this threshold can be improved through metacognitive training.

**Chapter 6 (Manuscript) - Computational Mechanisms of Detached Mindfulness**

Extends the models of metacognitive skill developed in Chapters 3 and 5 to the technique of detached mindfulness. It proposes computational mechanisms by which a detached stance toward affect reduces emotional reactivity, helping to explain its established clinical benefits.

**Chapter 7 - Discussion: AI and Future Directions**

Synthesizes the main theoretical claims of this dissertation, its limitations, testable predictions, and implications for artificial intelligence. It draws on the author's published work proposing that the theory set forth here can guide the design of metacognitively enabled AI systems. Finally, this chapter presents testable predictions for metacognitive skill across domains. It specifies concrete measures of metacognitive skill learning and outlines a study design and falsification criteria.

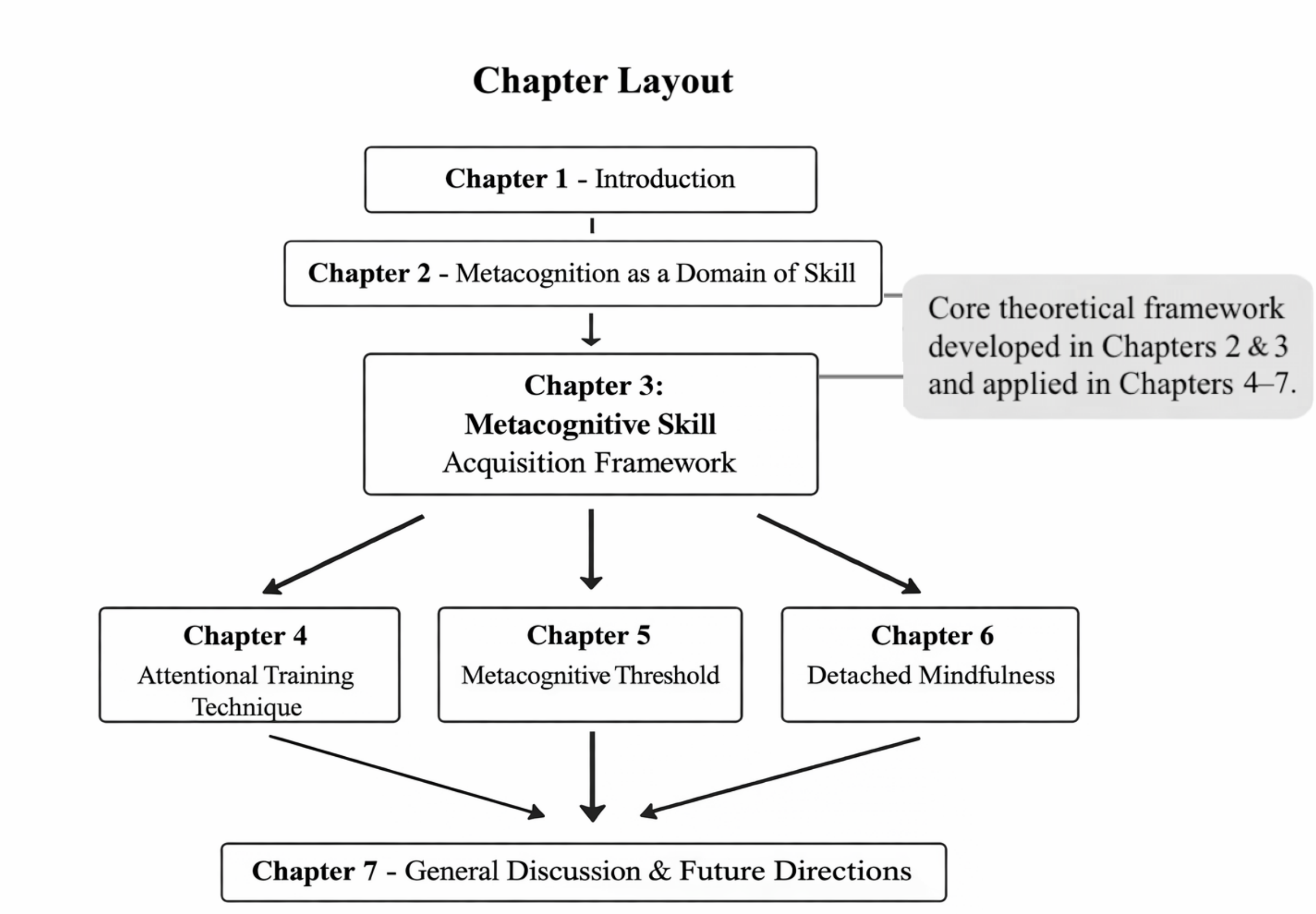

# Chapter 1. Introduction

This dissertation presents the first formal theory of metacognitive skill learning. It develops a mechanistic account of the cognitive and computational processes that enable humans to improve their ability to monitor and control their own mental processes. This theory explains a range of metacognitive phenomena and informs the design of metacognitively capable AI systems.

## 1.1 Background

Metacognition refers to the capacity to monitor and control one's own cognitive processes. The term was introduced by Flavell (1979) in his work on metamemory, building on earlier research on memory monitoring (Hart, 1965). Flavell applied the prefix *meta*, from the Greek μετά, meaning "beyond," to the word cognition to refer to self-directed mental processes, or "cognition about cognition."

Despite extensive research on metacognition, the field still lacks a mechanistic account of metacognitive learning. Here, metacognitive skill learning is defined as the process by which explicit monitoring and control strategies become more efficient, flexible, and automatic through practice. Decades of empirical work indicate that metacognitive performance can improve through deliberate practice, and that individual differences in metacognition are positively associated with performance in reasoning, attention, emotion regulation, learning, and problem-solving (Slagter et al., 2011; Keith & Frese, 2005; Veenman, 2015; Zawidzki, 2019). However, metacognitive theory has not kept pace with evidence that metacognition can be trained, and we still lack an account of how practice gives rise to more effective metacognitive control. As a result, the findings remain dispersed across studies, rather than integrated by a unifying framework. Basic questions remain unresolved, such as how explicit monitoring and control strategies become more efficient, flexible, and automatic through practice. This dissertation addresses this gap by proposing a formal, mechanistic account of metacognitive skill learning and applying it to specific metacognitive processes.

Metacognitive phenomena have long been discussed in Western philosophy under themes of self-knowledge, self-examination, and introspection, from the Delphic injunction to "know thyself" to Socrates' ideal of the examined life. The development of metacognitive processes in the form of explicit reasoning was largely directed toward epistemic goals such as greater certainty and truth. In Eastern traditions such as Buddhism, metacognitive skills take the form of meditation or mindfulness training, which involves the cultivation of attention and second-order attitudes (e.g., detachment). These metacognitive skills were primarily directed toward attention, emotion, and the reduction of psychological suffering. Together, these traditions highlight different families of metacognitive strategies, in the service of distinct metacognitive goals, that an account of metacognitive skill should be able to accommodate within a unified framework. This West/East contrast has significant overlap, however. Western philosophy includes metacognitively guided well-being aims (e.g., eudaimonia) pursued through self-reflection and reappraisal (e.g., Epictetus). Eastern traditions also contain epistemic rules emphasizing critical scrutiny and first-person verification (e.g., principles of Ehipassiko).

It is worth noting that the human capacity to make our own thoughts, emotions, and perceptions the object of focus has played a central role in scientific advancement. Scientific thinking itself has been described as "metacognitive in nature" (Kuhn & Pearsall, 2000) as it involves mental operations being explicitly directed at other mental operations, such as the prioritization of objective evidence and the development of explanatory theories. In this sense, the deliberate cultivation of metacognitive skill has been a major contributor to the accumulation of modern scientific knowledge. However, the capacity for cognition to turn inward has itself only recently become a direct object of scientific study. This is reflected in the growing attention to metacognition across cognitive science, psychology, education, philosophy, and artificial intelligence, where it is increasingly regarded as a phenomenon requiring explanation.

While metacognitive processes have been shown to be innate to humans, research demonstrates that they do not reach their full capacity without being deliberately cultivated. Despite decades of studies indicating that metacognitive abilities improve with training, only recently has metacognition begun to be thought of as a skill in its own right (Veenman, 2017; Lange, 2025a). Likewise, recent work has framed mental health as the exercise of metacognitive self-regulatory skill, not merely proper psychological functioning (Leder & Zawidzki, 2023). Despite this, a mechanistic account of metacognitive skill learning remains absent.

This issue extends to artificial systems. Unlike human cognition, metacognitive processes are not inherent to artificial intelligence and therefore must be engineered explicitly. Doing so requires a clear mechanistic account of how metacognition operates and how it can be learned. Metacognition is widely regarded as a major capability that is still lacking in contemporary AI (Johnson et al., 2026). This gap likely reflects, in part, an incomplete understanding of how metacognitive processes operate in human cognition, and how those processes can be instantiated in computational terms.

This dissertation addresses these issues by proposing a formal theory of the cognitive and computational mechanisms underlying human metacognitive skill learning. It does so by extending skill acquisition theories that have been successfully applied to motor and cognitive skills, namely the models advanced by Fitts (1964) and Anderson (1982, 2020). On this view, metacognitive skills can develop largely through proceduralization: the gradual transformation of explicit, declarative strategies into automatic, procedural routines for monitoring and control. This extension is novel, as metacognition has rarely been modeled as a learnable skill within a procedural learning architecture. Moreover, this extension of Anderson's (1982, 2020) computational account of cognitive skill learning provides a principled account of how metacognitive skills can be acquired and refined.

This theory is developed in Chapters 2 and 3, and applied across Chapters 4–7 to account for several metacognitive phenomena that have lacked mechanistic explanation, including attentional training, emotion regulation, the metacognitive threshold, and the therapeutic practice of detached mindfulness. Lastly, I consider how this computational account of metacognitive skill might be applied to AI capabilities, proposing its multiple realizability across both biological and artificial intelligence.

## 1.2 Theoretical Framing

This dissertation presents a unifying theoretical framework for metacognitive skill learning. Theoretical frameworks are essential for scientific progress as they unify empirical findings, guide the design of new studies, and sharpen the conceptual foundations of a field (Kuhn, 1962). A related methodological point was made by Newell (1973), who warned that without integrative theory, research can devolve into a long sequence of disparate experiments that fail to settle underlying issues.

This framing is also consistent with Brook's (2009) account of philosophy's contributions to cognitive science in three ways: i) improving the conceptual clarity and terminology of metacognition as a theoretical construct, ii) integrating a range of metacognitive data into a coherent theoretical picture, and iii) generating hypotheses for empirical testing.

The philosophical component of this thesis in Chapter 2 provides a conceptual analysis of metacognitive skill by drawing on accounts of skill in the philosophy of action. This chapter identifies domain-general features of motor and cognitive skill—such as goal-directed control, organization, knowledge, proceduralization, gradability, and constraints—and examines how these features apply to metacognition. This analysis will help to characterize the broader metacognitive phenomenon that later computational accounts will seek to explain.

### 1.3 Computational Modeling as Method

Computational modeling has a long history as a scientific tool for developing, testing, and refining theories of cognition (Newell & Simon, 1976; Marr, 1982). This methodology has developed alongside the growth of cognitive science as a field, as both are grounded in information-processing explanations of cognitive phenomena (Clark, 2013). Although this dissertation does not require a strong commitment to the Computational Theory of Mind (CTM), it assumes that computational frameworks can clarify cognitive theories by making explicit how information is represented and transformed (Fodor, 1975; Pylyshyn, 1984).

This dissertation adopts Dennett's (1978) proposal that computational modeling can serve as a disciplined tool of philosophical investigation. Dennett argues that,

> It is notoriously difficult to keep wishful thinking out of one's thought-experiments; computer simulation forces one to recognize all the costs of one's imagined design. As Pylyshyn observes, 'What is needed is … a technical language with which to discipline one's imagination.' The discipline provided by computers is undeniable. (p. 117)

Dennett's insight is that formalizing a theory within a computational framework helps to constrain speculation, as it requires mechanisms to be explicitly specified. Rather than relying on verbal definitions alone, computational modeling forces a theorist to confront real issues of implementation. In this way, computational models function not only as tools for prediction but as methods for refining and evaluating theoretical claims about cognition.

Cognitive modeling is especially valuable when implemented within a biologically grounded cognitive architecture that is constrained by neural and behavioral data (Stocco et al., 2021), as in the present dissertation. In this context, models aim to provide mechanistic explanations of cognitive processes that are consistent with how the brain operates. This approach was largely motivated by Newell's (1990) call to develop a unified theory of cognition, which would specify a single set of mechanisms that give rise to the full range of psychological processes and behavior. The distinction between cognitive models and cognitive architectures is important because they play different roles. A cognitive model specifies a particular psychological process in runnable form, whereas a cognitive architecture provides the larger mechanistic framework for how multiple cognitive processes operate in a unified system (Sun, 2006). This distinction is important for metacognition, which depends on interactions among monitoring, control, memory, attention, knowledge, and action selection.

Within Chapters 3–6, the proposed theory is implemented within a biologically grounded cognitive architecture, allowing hypotheses about metacognitive skill to be specified and evaluated in mechanistic terms. By doing so, this account provides a more precise explanation of how metacognitive processes operate and are acquired. Exploratory computational models of metacognitive monitoring and control developed for this work are available in a public GitHub repository (Conway-Smith, 2026).

Chapter 2 will first draw on work in the philosophy of cognition and action to characterize metacognition as a domain of skill. This provides the conceptual groundwork for the computational accounts that will be developed in later chapters.

# Chapter 2. Metacognition as a Domain of Skill

This chapter lays the philosophical and conceptual foundations for regarding metacognition as a genuine domain of skill in its own right. This framing situates metacognitive skill within the broader literature on motor and cognitive skills, highlighting shared properties such as goal-directed action, hierarchical structure, the interplay of declarative and procedural knowledge, and the gradual automatization of control with practice. By bridging the theoretical and empirical work on skill acquisition with existing research on metacognition, this chapter specifies a set of domain-level features that any robust theory of metacognitive skill learning must accommodate.

The material in this chapter is adapted from Conway-Smith (2025), "Metacognition as a Domain of Skill," published in the *Proceedings of the Annual Conference of the Cognitive Science Society* (Vol. 47).

## 2.1 Introduction

Interest in metacognition has grown alongside the emergence of information-processing models of human cognition. Metacognition, the monitoring and control of cognitive processes (Flavell, 1979; Proust, 2013), continues to be a prominent area of research in cognitive science within fields such as psychology, philosophy, education, and artificial intelligence. While metacognition has been shown to be an improvable skill, its characterization as a domain of skill remains underdeveloped. To help articulate metacognition as a distinct domain of expertise, this chapter situates metacognitive skill alongside traditional domains of skill, drawing on research that identifies their shared characteristics (Shepherd, 2021) as well as dual-system theories of metacognition (Thompson, 2009). Here, I argue that metacognitive expertise shares core principles with other skill domains while pursuing distinct goals, such as the regulation of attention, emotion, and memory.

*First*, I outline key characteristics shared by different skill domains, including control, goal hierarchies, knowledge, and action restrictions. *Second,* metacognitive skill is positioned within this framework, demonstrating its alignment with existing models while highlighting its unique properties. *Third,* the gradability of metacognitive skill is examined across success rates, goal breadth, and adaptability.

## 2.2 Characteristics of Skill Domains

This section will outline the fundamental attributes that define skill domains across diverse fields. Whether it be driving, chess, or attentional control, skill domains share core components, including specialized knowledge, hierarchical goal structures, and constraints on both action-types and applicable circumstances. Identifying these shared characteristics provides a conceptual framework for positioning metacognition within the broader landscape of skill, which enables a more precise articulation of its distinct actions, contexts, and objectives.

This discussion is drawn in large part from the research of Joshua Shepherd (2021) and his work on action domains, which begins by emphasizing control as a fundamental component of any skill. In this view, one cannot exercise a skill S without possessing control over behaviors involved in exercising S. Researchers widely agree that skilled action requires agents to possess high levels of control over their activity within a domain. This, in turn, often requires years of deliberate practice.

### 2.2.1 Ideals, Goals and Actions

Shepherd characterizes control as an agent's ability to flexibly and consistently align their behavior with a planned course of action, which is measured by an ideal of success. An ideal qualifies actions based on their outcomes or overarching goal, and serves as a fundamental element of any action domain. Control, in this sense, depends on causal factors that enable an agent to reliably execute actions in alignment with their ideal or overarching goals and the plans to achieve them. Some action domains have only one ideal of success, while others consist of more than one ideal. For instance, the overarching ideal or goal in chess is to checkmate one's opponent, while gymnastics involves a combination of ideals such as kinesthetics, balance, and form.

Goals often require subgoals to achieve, and entail a hierarchical goal structure where goals and tasks are organized according to their conduciveness to the ideal of success. For instance, in basketball, the goal of scoring the most points requires subgoals, such as shooting accurately and defending against opponents' attempts to score. These in turn are supported by further subgoals such as ball handling and footwork. Subgoals can be ordered according to their importance or centrality to higher goals, that is, according to their conduciveness to success. While some subgoals are critical to success, others are peripheral, and their contributions are minor. Specific goals and subgoals within a domain require particular action-types for their achievement. The appropriate behaviors and action-types possess causal properties that consistently lead to the attainment of goals.

Expertise typically involves agents being skilled at more than one specific action and instead requires proficiency in a cluster of action-types. These clusters tend to support each other both heterarchically and hierarchically. Skills that are clustered heterarchically occur simultaneously. For instance, a tennis player's forehand serve is improved when combined with a particular body position. Action-types may also be organized hierarchically, or linearly, such as a basketball player's skillful dribbling toward the basket supporting an eventual layup. In the case of cognitive skills such as math, basic addition and division are required to solve more complex equations. The subgoals and task structures of motor and cognitive domains of skill depend on their conduciveness to their ideals of success or overarching goal. Certain action-types, like aiming, are shared across multiple domains, such as archery and football, demonstrating a partial fluidity between domains.

Restrictions help to define and regulate skill development. Action-type restrictions limit permissible behaviors within a domain, as seen in sports rules (e.g., soccer players cannot use their hands) or professional guidelines (e.g., medical professionals must follow ethical protocols). Circumstance-type restrictions constrain where and when skills can be applied, such as playing fields in sports or syntax rules in programming. These restrictions define the space of permissible action within a domain.

### 2.2.2 Knowledge

Research on skill has long been interested in the types of knowledge that skilled experts rely on. Classically, this issue has been framed in terms of the distinction between *knowing that* and *knowing how* (Ryle, 1949), and later recast in cognitive science as a distinction between declarative and procedural knowledge (Cohen & Squire, 1980; Dreyfus & Dreyfus, 1986).

*Declarative knowledge* is formatted propositionally, and encompasses explicit facts, rules, and strategies relevant to a domain. For instance, building a fire requires knowledge of the step-by-step procedures, while chess players must understand the appropriate rules about moves. Experts possess robust internal models of their domain, which are crucial for controlling complex actions. While not always necessary, domain-specific knowledge enhances planning and execution. For example, chess experts rely on knowledge of goals and action-types, such as piece movements and strategic deployment (de Groot, 1978). Internal models represent causal relationships within a domain, allowing agents to predict outcomes and select appropriate actions. A driver, for instance, relies on an internal model of surrounding vehicles, adjusting their behavior based on changing characteristics like speed, size, and maneuverability. These models support the selection of success-conducive plans, i.e., mental representations outlining sequences of actions directed toward a goal. For instance, fire-building techniques range in their effectiveness, and one's expertise involves refining the most efficient approach. While explicit knowledge enables an individual to articulate conditions for success and the strategies needed to achieve goals, it does not necessarily execute the actions themselves.

*Procedural knowledge*, by contrast, is implicit and often non-verbal, relating to the specific process of executing tasks within a domain. It encompasses motor and procedural representations that direct and control actions that reliably result in the attainment of goals. For example, a tennis player may use declarative knowledge to learn proper serving techniques, while procedural knowledge enables the player to physically execute the serve with timing and precision. Through practice, procedural knowledge is developed and refined to enable actions to become increasingly efficient, fluid, and automatic.

### 2.2.3 Proceduralization

A significant degree of skill acquisition involves a process of automatization, where an agent's actions transition from deliberate control to more automatic execution (Fitts, 1964; Logan, 1988; Anderson, 1982, 2020). At the core of proceduralization is the shift from declarative to procedural knowledge. Early in learning, individuals rely heavily on declarative, explicitly accessible facts and rules to consciously guide performance. Through repeated practice, this slow and effortful retrieval process is progressively replaced by procedural knowledge, which enables faster and more efficient task execution with minimal cognitive effort (Fitts & Posner, 1967; Anderson, 1982;

Kim & Ritter, 2015). This process enhances fluency and adaptability, enabling experts to operate effectively in dynamic and unpredictable contexts. Importantly, automatization does not eliminate cognitive control entirely. Instead, routine performance becomes automatic, while higher-order processes are reserved for error detection, strategy revision, and goal adjustment (Fridland, 2019).

While some domains require the automatization of a narrow range of specialized skills (e.g., a baseball pitcher refining throwing mechanics), others require the integration of a broader set of skills (e.g., a trial lawyer combining reasoning, persuasion, and speaking). In both cases, skill is gradable, varying in proficiency, success rate, task range, and adaptability. To possess partial skills means an agent excels in certain dimensions while being less developed in others, which highlights another important aspect of skill—its gradability.

### 2.2.4 Gradability of Skill

The gradability of skill means that an agent can possess varying levels of proficiency within a domain. Shepherd (2021) proposes three principal dimensions along which a skill may vary (Figure 2.1): the success-rate at achieving goals (height), the range of goals achieved (breadth), and performance across diverse circumstances (depth).

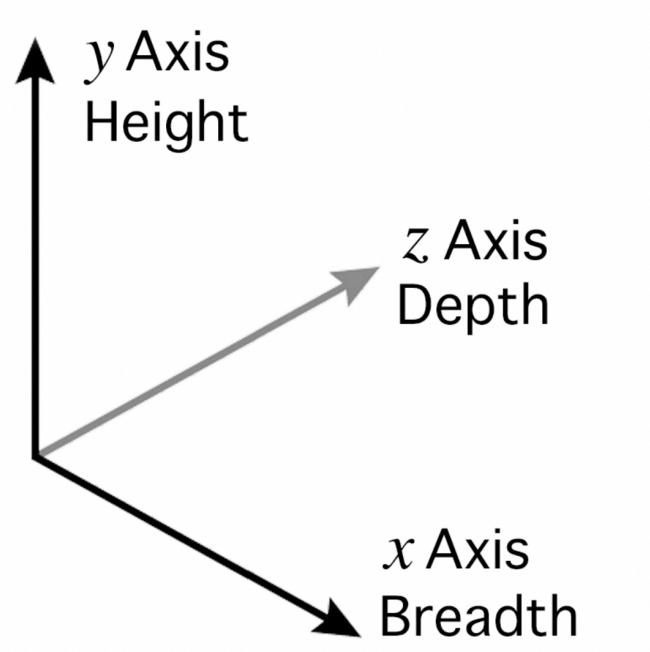

**Figure 2.1.** Three dimensions of skilled performance.
The origin (0,0,0) represents novice-level performance.

The dimension of *height* refers to the agent's actual success rates at goals that are central to a domain. A higher success rate in achieving these goals typically reflects greater skill. In practice, priority is often given to actions that contribute most to overall success, or that advance the most central goals within the domain. The dimension of *breadth* involves the agent's success rate across the range of goals within a domain. An agent is considered more skilled if they can maintain strong performance across a wider set of action-types or sub-skills, again with greater weight placed on the domain's more central goals. The dimension of *depth* evaluates the range of circumstances under which an agent can maintain good performance. This captures the flexibility of skill, as distinct from the brittleness of habit: skilled control can adapt to novel situations, such as golfers adjusting their swings to changing wind conditions and terrain. The more varied the situations in which an agent can sustain strong height and breadth, the greater their degree of skill.

These three dimensions interact and contribute to a more nuanced understanding of skill, with the ideal being high performance across all dimensions. This provides insight into how an agent exerts control to align their actions flexibly and consistently with a goal-oriented plan, whether motor, cognitive, or metacognitive.

### 2.3 Metacognition

Flavell (1979) situated metacognition within a broader account of cognitive monitoring and regulation. He emphasized the interplay among metacognitive knowledge, tasks, and strategies, which later work has summarized as both the monitoring and control of cognitive operations (Proust, 2019; Shea et al., 2014; Wells, 2019). *Metacognitive monitoring* is the capacity to perceive and identify cognitive states in ways that can guide regulation or behavior. This can include awareness of one's thoughts and epistemic feelings, such as a feeling of knowing or uncertainty.

*Metacognitive control* refers to actively regulating cognitive states or processes in the service of explicit metacognitive goals.

The interaction between these processes allows for reciprocal adjustments, enabling individuals to monitor and control various cognitive processes like attention, emotions, learning, and memory. For example, one may notice their attention drifting while driving and use metacognitive control to refocus. In the same situation, one might identify a distracting emotion and self-regulate through calming techniques. Similarly, a new driver may struggle to remember the correct sequence of actions for changing lanes and employ a metamemory strategy such as an acronym to help with recall.

Metacognitive skill refers to the extent to which one can monitor and control their own cognitive processes (Van der Stel & Veenman, 2010). These skills represent higher-order cognitive abilities that regulate thoughts, emotions, and mental processes. Empirical research suggests that metacognitive ability can be improved through practice and training (Meher, Baral, & Bhuyan, 2021; Schaeffner et al., 2021). For example, educational interventions can enhance problem-solving, self-regulation, and academic performance (Zimmerman & Schunk, 2011), while attentional processes can be strengthened through repeated practice (Posner et al., 2015). Metacognitive training, including mindfulness techniques, can also improve cognitive control and emotion regulation, as demonstrated in Cognitive Behavior Therapy (CBT; Dobson, 2013) and Metacognitive Therapy (MCT; Normann & Morina, 2018), both of which support the regulation of maladaptive thoughts and emotions (Wells, 2019). Various psychiatric disorders, such as anxiety and depression, have been shown to reflect deficits in metacognitive skill, making its development a promising therapeutic approach that is distinct from purely pharmacological or discursive methods (Leder & Zawidzki, 2023).

### 2.3.1 Metacognition as a Skill Domain

This section connects the previously discussed domain-general characteristics of skill to the specific properties of metacognition. While the literature covers a broad range of metacognitive abilities including metamemory, metareasoning, and metalearning, this discussion focuses on paradigmatic examples of metacognitive skill in attentional control and emotion regulation.

### 2.3.2 Metacognitive Goals

This section seeks to clarify the underexplored nature of metacognitive goals for the purpose of better defining metacognitive skill as the reliable attainment of such goals. Expert metacognitive control requires the ability to flexibly and reliably align *mental actions* with a plan, evaluated against an ideal of success. Here, an ideal represents an overarching standard for an optimal outcome in a specific metacognitive subdomain (e.g., an ideal attentional, emotional, or learning outcome). Within the literature on attentional training, various ideals of success are studied, ranging from achieving a maximally stable narrow focus to all-inclusive open monitoring (Lutz et al., 2008; Eberth et al., 2019). In this context, metacognitive control involves continuously monitoring and adjusting one's mental actions to align with overarching goals and subgoals.

Within the literature on metacognition, there has been remarkably little investigation into the precise nature of metacognitive goals. The specific properties of metacognitive goals are largely left implicit and treated as a given. The bulk of the research instead focuses on characterizing metacognitive processes, for example by contrasting meta-level vs. object-level, procedural vs. analytic, conceptual vs. non-conceptual, implicit vs. explicit, model-based vs. model-free, System/Type 1 vs. System/Type 2, or attributivist vs. evaluativist accounts (Nelson & Narens, 1990; Shea, 2024b; Carruthers & Williams, 2022; Proust, 2013).

Within the computational cognitive science literature, metacognitive goals are usually treated functionally rather than as explicit, propositionally structured states. In metareasoning and resource-rational models, they typically appear as objective functions or utility criteria that specify what counts as good performance for meta-level control, for example maximizing expected reward given the costs of computation (Lieder et al., 2018). Metacognitive goals have also been treated as constraints over object-level reasoning and planning, without close analysis of their representational format. In a problem-solving context, Elliott et al. (2019) state that metacognitive goals are the objectives to be attained in a problem space, with metacognitive actions defined as the operations directed at attaining those goals.

Articulating the properties of metacognitive goals is crucial for differentiating metacognitive control from motor or cognitive control. Some have distinguished metacognitive control as a form of control that directly targets mental processes, as opposed to other forms of control that merely make use of mental processes (Lange, 2025a). For example, a tennis or chess player controls their cognition instrumentally to succeed in performing an external motor movement or strategic maneuver.

Lange (2025a) differentiates these forms of control according to the content of their respective goals. On the one hand, *cognitive control* regulates cognitive processes instrumentally, as a means of attaining an externally directed goal. On the other hand, *metacognitive control* occurs when the control targets mental processes or states directly, as ends in themselves. Lange's novel proposal is to define metacognitive control in terms of the *metacognitive goals* that the control processes are directed toward. Lange illustrates this by way of an example wherein Lucy controls her cognitive processes in her attempt to find a coffee shop. While the goal of finding a coffee shop does shape Lucy's thinking and motor activity, the content of this goal does not concern these mental processes—it concerns the external environment and her position within it. If Lucy were to direct her control processes toward mindfully attending to her breath, with the intention of stabilizing her first-order attention as an end in itself, then she has implemented a metacognitive goal. It is important to note that Lange's stated aim was to provide an account of the form of metacognitive control specific to mindfulness practice, rather than metacognitive control more broadly.

This section will incorporate several of the attributes that Lange commits to in his 2025 paper, in which he distinguishes metacognitive goals by their content. On his view, *metacognitive goals* have the following properties:

i) Metacognitive goals are *goals about one's own mental processes.*

ii) They can be accessed and reported by the agent in *propositional format.* Although there is literature on implicit or procedural metacognition, Lange does not commit to non-propositional metacognitive goals. He regards the difference between implicit metacognitive goals and instances of cognitive control as a line too finely drawn, and outside the scope of his paper.

iii) They are *personal-level goals* (which agents can access and report), as opposed to sub-personal-level goals (which agents cannot access or report). Lange also refers to these as *proximal goals*, concerned with the here-and-now regulation of one's own mental processes. Importantly, proximal goals can be coupled with ultimate goals. For instance, Lucy can have a proximal metacognitive goal of focusing her attention on her breath, in service of the ultimate goal of stabilizing her attention or achieving a therapeutic effect.

iv) Within Lange's mindfulness framing, metacognitive goals can involve *first-order attentional and second-order attitudinal* aims, specifying both the focus of attention and the stance one takes toward one's own mental states. *First-order attentional* goals are personal-level metacognitive goals whose content concerns directing first-order attention toward objects or processes that do not themselves constitute the agent's attention (such as the breath or bodily sensations). *Second-order attitudinal* goals are personal-level metacognitive goals whose content concerns the attitudes an agent adopts toward her own occurrent mental states and processes, such as remaining non-reactive to passing thoughts. Hence, Lucy's focused attention meditation can be understood as implementing first-order attentional goals, while her open monitoring practice implements second-order attitudinal goals that target her own first-order processes. While some metacognitive ideals allow for a plurality of goals, others may encompass only a single goal. For example, achieving a maximally stable focus may involve training attention on a specific sensory modality, such as sight, sound, or the breath.

The attainment of metacognitive goals is supported by subgoals and their corresponding action-types. Typically, metacognitive subgoals that best support the ideal of success are prioritized. In developing stable narrow focus, subgoals may include ignoring irrelevant stimuli and preventing mind-wandering (Fox et al., 2016). Subgoals can be trained independently, with action-types applied both hierarchically (e.g., clearing the mind before focusing) and heterarchically (e.g., focusing while simultaneously resisting mind-wandering). Emotion regulation similarly involves controlling mental actions to achieve an ideal state, such as one entirely free from anxiety (Gross, 2014). Various approaches can be used to achieve these ideals, including cognitive reappraisal, attentional training, and mindfulness techniques (Gross, 2015; Dobson, 2013; Wells, 2019). In many cases, effective emotion regulation involves polyregulation, that is, the coordinated use of multiple regulatory strategies within the same situation in order to respond more adaptively to shifting demands (Ford, Gross, & Gruber, 2019). For instance, in managing anxiety, subgoals may include identifying emotional triggers, disidentification, and reducing reactivity to them. These subgoals can be trained independently through additional subgoals and action-types, while others may transfer from other metacognitive domains like attentional training. Proceduralization likely plays an important role in polyregulation by automatizing individual strategies, thereby offloading them from working memory and allowing them to operate in tandem.

From this, it follows that metacognitive skills in attentional control and emotion regulation encompass a diverse range of goals, subgoals, and mental action-types. Similar to motor and cognitive skills, metacognitive skill relies on factors that enable reliable control of mental actions, aligning them with plans, goals, and ideals of success.

### 2.3.3 Metacognitive Knowledge

Skilled metacognitive control largely depends on domain-specific metacognitive knowledge—metarepresentations—a propositional form of declarative knowledge that refers to cognitive or metacognitive properties (Shea et al., 2014; Proust, 2013). Metarepresentations help to build internal models that capture the key attributes, processes, and causal relationships within a metacognitive subdomain. This aligns with the distinction some researchers have made between model-based and model-free metacognition (Carruthers & Williams, 2022). *Model-based metacognition* corresponds to Type 2 or analytic metacognition, as it involves explicit metarepresentations that form organized theory-like, causal structures of the domain. In this case, mental actions are selected by searching through and evaluating options in the model. In contrast, *model-free metacognition* corresponds to Type 1 or procedural metacognition, involving single implicit signals that guide downstream first-order processes. In this case, mental actions occur in the absence of an explicit theory of one's own mind or concept-like representations. Mental actions are generated by retrieving cached values of those actions (acquired through associative learning) and are selected automatically, and independently of changes in the environment.

For our purposes, model-based metacognition is taken to be essential for metacognitive skill, since it enables agents to predict, deliberately select, and adjust the mental actions needed to achieve their goals. For example, skilled attentional control depends on internal causal models that represent the target of focus, the actions required to sustain attention, and potential obstacles such as intrusive thoughts (Jahn et al., 2023). Emotion regulation can likewise be supported by causal knowledge of monitoring and control strategies that reliably reduce rumination and affective disruption (Wells, 2019). Similarly, explicit metamemory strategies allow agents to intervene causally to improve memory storage and recall while reducing the likelihood of forgetting (Dunlosky et al., 2013). By drawing on internal models, individuals can flexibly select and adapt their strategies to respond to changing demands. Thus, model-based, or Type 2, metacognition involves the kind of explicit conceptual knowledge required for genuinely agentive action in the strongest cases of metacognitive control.

Importantly, metacognitive knowledge differs from metacognitive skill, as knowledge does not necessarily lead to the automatic deployment of metacognitive processes (Veenman & Elshout, 1999). In all action domains, the execution of declarative instructions relies on procedural knowledge to carry out the task. Similarly, through repeated practice of metacognitive instructions, domain-specific and task-specific procedural knowledge is developed and refined into more skillful forms. Veenman et al. (2006) argue that metacognitive skills are best understood as domains of procedural knowledge. For example, attentional training techniques may begin as declarative instructions and, through practice, develop into the procedural knowledge required for fluid, automatic execution (Anderson, 2016). This is consistent with perspectives that treat metacognition as a form of skilled know-how, akin to physical and cognitive skills, yet focused on the control of one's own attention and emotion (Zawidzki, 2019).

### 2.3.4 Procedural Metacognition and Its Limits

While procedural knowledge is crucial to the deployment of skillful metacognitive control, I argue that procedural (i.e., model-free, Type 1) metacognition does not qualify as a robust example of explicit metacognitive control, particularly when control processes are driven by metacognitive feelings. Procedural metacognition consists of affect-based, non-conceptual mechanisms that monitor and regulate cognitive actions such as memory retrieval, perceptual discrimination, and problem solving (Proust, 2019; West & Conway-Smith, 2019). On this view, procedurally-driven metacognitive control is motivated by non-conceptual signals or experiences (e.g., feelings of knowing), rather than through explicit metacognitive representations (Fleming et al., 2012; Shea et al., 2014). This is in contrast to analytic (i.e., model-based, Type 2) metacognition, which involves conceptually rich and verbally articulable representations of one's own mental states and processes.

Procedural metacognition can include epistemic (noetic) feelings that guide memory search and metareasoning, such as feelings of knowing (FoK), confidence, and fluency (Koriat, 2000; Ackerman & Thompson, 2017). In cases such as these, the aim of remembering, or arriving at a correct answer, does not involve an explicitly represented metacognitive goal. Rather, memory recall is part of the base cognitive architecture, where an external cue triggers guidance processes that unfold without being indexed to any explicitly represented goal. While metacognitive feelings can be explicit in their own right and consciously available, they are not propositionally structured, and function more like analog signals than sentence-like representations. Typically, we do not intend to generate feelings of knowing or fluency, nor are these states propositionally formatted, in the sense that they are not truth-evaluable mental representations composed of concepts. Metacognitive feelings are better understood as evaluative signals that can be used for implicit guidance, rather than as propositional states.

While Type 1 metacognitive feelings can be adaptively useful, they can also be shaped by affective and cognitive biases. Feelings of knowing and confidence may be distorted by emotional reasoning, processing fluency, confirmation bias, and the illusory truth effect, where individuals prioritize subjective feelings of truth over objective evidence (Alter & Oppenheimer, 2009; Fazio et al., 2015). In such cases, agents often default to a heuristic that treats the phenomenology of certainty as a reliable signal of truth (i.e., "I feel that I know, therefore I know"). Shea's (2024a) account of a "feeling of reliability" makes a similar point: inferential reasoning can be accompanied by a procedural sense that one's reasoning is trustworthy, even when the agent lacks an explicit justification.

The tendency to mistake feelings of certainty for genuine certainty may be described here as *naïve metacognition*. By analogy with naïve realism, which treats perception as an accurate presentation of the world, naïve metacognition treats feelings of certainty as accurate signals rather than as fallible internal cues. By contrast, Type 2 metacognition allows for explicit representation and evaluation of Type 1 biases, enabling agents to recognize that their feelings of knowing do not necessarily track truth or competence (Fleming & Lau, 2014). Through explicit monitoring and deliberate metacognitive control, individuals can better avoid biases and errors that might otherwise go unnoticed. In this way, common errors associated with automatic Type 1 biases can be mitigated through the cultivation of deliberate Type 2 metacognitive skill.

Type 1 and Type 2 metacognition are guided by different epistemic cues. That is, each system is sensitive to different signals that help determine whether a mental action "should" take place. Proust (2013) describes this as a form of normative governance: metacognitive processes regulate cognition in relation to norms or standards for successful cognitive action. Type 2 metacognition is primarily sensitive to epistemic cues appropriate to propositional knowledge, such as logic, evidence, and coherence. By contrast, Type 1 metacognition is primarily sensitive to cues associated with epistemic feelings, such as feelings of knowing, fluency, and confidence.

Metacognitive beliefs can play an important role in motivating which epistemic cues an agent prioritizes when directing their mental actions. Metacognitive beliefs are beliefs about one's own thinking and about how cognitive processes should be monitored and controlled (Wells, 2009). For example, a person who believes that rational evidence is a more reliable epistemic signal than feelings of knowing will prioritize evidentiary cues when regulating their mental actions. Conversely, an individual may view their epistemic feelings as trustworthy and therefore privilege "intuition" while discounting rational evidence. In this way, metacognitive beliefs help determine which cues an agent treats as authoritative in guiding their own cognition. Because Type 1 epistemic feelings often function automatically to bias judgment, prioritizing Type 2 evidential cues typically requires supportive metacognitive beliefs as well as sustained training and practice. In this sense, rational or scientific thinking can itself be understood as a form of metacognitive skill that must be deliberately cultivated.

Importantly, this does not mean Type 1 epistemic feelings are always unreliable. Rather, their reliability depends on the extent to which they have been calibrated through experience. In novices, feelings of knowing or certainty may be only weakly connected to actual competence, and may produce overconfidence. In experts, however, deliberate practice can refine intuitive judgments so they reflect meaningful patterns in a domain such as chess, allowing them to guide highly skilled performance (Kahneman & Klein, 2009). In this sense, expert intuition is not opposed to evidence or training, but often results from the proceduralization of practice.

Metacognitive skill, in the most robust cases, involves processes that are deliberate, effortful,

and indexed to an explicit metacognitive goal, distinguishing it from more procedural forms of metacognition. There is some evidence, for instance from perceptual decision-making tasks, that individuals can improve the correspondence between their confidence judgments and their actual competence (Carpenter et al., 2019). However, we consider becoming more proficient in some forms of procedural metacognition by way of increased experience to be categorically different from metacognitive skill. Moderate improvements acquired by happenstance, association, or even indirect intention do not equate to the kind of explicit, intentional metacognitive expertise that is our primary focus.

### 2.3.5 Partial Skills

In some metacognitive domains, agents may require only partial skills to succeed, while others necessitate a broader range of metacognitive strategies. For example, achieving single-pointed focus may demand mere resistance to mind-wandering, while maintaining broader attentional goals (such as open monitoring, or impermanence perception) may require a cluster of skills, including metacognitive sensitivity, adaptability, and the ability to shift focus flexibly. To possess partial metacognitive skills means to excel in certain subdomains while being less developed in others, underscoring the gradability of metacognitive expertise.

### 2.3.6 Gradability

Metacognitive skills are gradable, meaning that monitoring and control processes can vary in proficiency. Here, we examine the gradability of attention and emotion regulation across three key dimensions—height (success rate), breadth (range of goals achieved), and depth (adaptability)—extending Shepherd's (2021) three-dimensional framework to the domain of metacognition.

*Height* in the context of metacognitive skill refers to an individual's success rate in achieving metacognitive goals. Higher proficiency in sustaining attention, for example, indicates greater effectiveness in this form of metacognitive control. Priority is generally given to mental actions that most effectively contribute to this central goal. In attentional training, concentration on a target of focus takes precedence over peripheral subgoals like clearing the mind. In emotion regulation, sustained attention may serve as a secondary goal, supporting broader objectives such as attaining a stable perception of affective impermanence for the purpose of emotional detachment.

*Breadth* captures an agent's success across various metacognitive goals and subgoals. Greater attentional skill is expressed as proficiency across multiple action-types. In the case of attentional control, this can include resisting intrusive thoughts, open monitoring, and narrow focus. Similarly, emotion regulation skills that enhance one's perception of affective impermanence rely on several attentional capacities, including focused attention, open awareness, metacognitive sensitivity, and non-reactive observation.

*Depth* reflects the ability to flexibly maintain metacognitive control across diverse and dynamic circumstances. It measures how well an agent can adapt their mental actions to novel or unexpected challenges. For example, an individual with novice-level depth of attentional control may only be capable of focusing on a single attentional task, and only in low-stimulus environments. Conversely, an expert may be able to control their attention flexibly and reliably across diverse tasks and high-stimulus environments. Similarly, a novice in emotion regulation may only be able to regulate a single emotion in particular circumstances, whereas an expert could regulate a range of emotions across a variety of unexpected situations.

To make these dimensions more concrete, consider how they may unfold in emotion regulation training. In this scenario, the skill development of a novice helps to illustrate how height, breadth, and depth can change with practice. Initially, the novice exhibits limited height (low success rates), narrow breadth (few skills), and shallow depth (limited adaptability to intense or changing emotional states). Novices often begin by applying a domain-specific metacognitive plan (e.g., non-reactive open monitoring), typically communicated by an instructor, therapist, or another form of structured guidance. Through repeated practice of explicit instructions, the practitioner builds up procedural knowledge that supports a higher success rate for longer periods, greater breadth in goal attainment (e.g., resisting mind wandering and maintaining detachment), and deeper flexibility across circumstances (e.g., applying strategies to both sadness and anger, at home and at work). The specific details by which metacognitive skill becomes proceduralized are explained in Chapter 3.

### 2.4 Restrictions

Restrictions on metacognitive action-types and the contexts in which they are applied influence both the development and practical use of metacognitive skills.

*Action-type Restrictions.* Metacognitive skill is inherently constrained by the types of mental actions an agent can perform. These boundaries define what is possible within metacognitive domains and guide skill refinement. For instance, equanimity, an effective emotion regulation strategy, requires focusing on the impermanence of affective sensations rather than visual or auditory stimuli (Wongpakaran et al., 2021). In this case, the target of focus is critical to achieving desired metacognitive outcomes. Certain restrictions also differentiate legitimate metacognitive actions from external enhancements. For example, while caffeine may temporarily improve attention, as an external supplement it does not constitute a metacognitive action. Likewise, metacognitive expertise excludes pharmacological interventions, as they bypass skill development and do not produce enduring improvements in deliberate, self-directed metacognitive processes.

*Circumstance-type Restrictions.* The effectiveness of metacognitive training is also shaped by environmental and situational constraints. Novices require controlled, low-distraction settings—silence, dim lighting, or the absence of emotionally charged stimuli—to develop attentional focus and emotion regulation. In contrast, experts can sustain metacognitive control in more dynamic and unpredictable environments, demonstrating greater adaptability and depth of expertise. These restrictions evolve alongside skill development. A novice meditator may struggle to maintain focus in a noisy environment, while an expert can sustain equanimity even within relative chaos, such as during emotionally charged interactions or high-stakes situations. This underscores the importance of tailoring metacognitive training to match the context and an individual's level of expertise.

## 2.5 Conclusion

This chapter lays out a case for metacognition as a distinct skill domain that shares core principles with motor and cognitive skills. Framed this way, metacognitive expertise involves structured, goal-directed mental action, organized around ideals of success and supported by domain-specific knowledge and control. On this view, metacognitive skill involves not merely possessing metacognitive knowledge or applying the occasional strategy but the ability to reliably and flexibly align mental actions with a plan in pursuit of a metacognitive goal.

This account also clarifies why not every improvement in metacognitive performance amounts to metacognitive skill. Procedural metacognition and metacognitive feelings may guide cognition adaptively, but the strongest cases of metacognitive skill involve deliberate, flexible, and goal-indexed control. Finally, like other skill domains, metacognitive expertise is gradable: agents may vary in their success rate, range of metacognitive goals, and the ability to adapt across varied and changing circumstances.

# Chapter 3. Metacognitive Skill: How It Is Acquired

This chapter presents the dissertation's core theoretical proposal by developing an account of the cognitive and computational mechanisms that underlie metacognitive skill acquisition. It advances the claim that metacognitive expertise develops through proceduralization within the Fitts–Anderson skill acquisition framework (Fitts, 1964; Anderson, 1982, 2020). This framework of metacognitive proceduralization constitutes a novel extension of a model that has been effectively applied to both motor and cognitive skill acquisition. It clarifies the mechanisms through which practice and training enhance metacognitive control processes, and specifies the modules and information types involved in their operation. As a result, the framework helps explain a range of otherwise puzzling empirical findings across domains. This chapter provides the foundational theory that will be applied to the empirical and computational work developed in Chapters 4–7.

The material in this chapter is adapted from Conway-Smith, West, and Mylopoulos (2023), "Metacognitive skill: How it is acquired," published in the *Proceedings of the Annual Conference of the Cognitive Science Society*.

## 3.1 Introduction

Metacognition is the monitoring and control of cognitive processes, and it has received increasing attention in recent years across numerous fields. While metacognition has been shown to be a learnable skill across domains such as attentional training, emotion regulation, and metamemory (Posner et al., 2015; Shin et al., 2015; Normann & Morina, 2018; Carpenter et al., 2019), the mechanisms that give rise to metacognitive skill remain unclear. This theoretical gap poses a barrier to metacognition research and may impede its application in educational and clinical contexts. It is often noted that metacognition, as a theoretical domain, "lacks coherence," and that more work is needed to specify its underlying mechanisms (Veenman et al., 2006; Dunlosky & Rawson, 2019).

We propose that a coherent and parsimonious framework for understanding metacognition can be derived by situating it within Fitts' (1964) classic skill acquisition framework, as computationally interpreted in ACT-R (Anderson, 1982, 2020). To date, a range of metacognitive theories have been modeled within the ACT-R cognitive architecture. These include metacognition instantiated as a separate metacognitive module (Anderson & Fincham, 2014), recursive self-processing (Kralik et al., 2018), reflection driven by affective signals (Juvina et al., 2018), and as reinforcement-based evaluation (Wu et al., 2025). The approach developed here differs in that it sets forth a novel theory that bridges metacognition research with the skill acquisition literature to clarify the mechanisms underlying metacognitive skill learning.

*First,* we offer a brief review of the relevant literature on both metacognition and skill theories. *Second*, we explain how the classic skill acquisition models of Fitts and Anderson can be applied to the domain of metacognition. *Third*, we will discuss how the proposed framework sheds new light on the nature of metacognitive skill, and how this helps to explain otherwise unexplainable data within the literature.

The skill acquisition theories relied on here largely involve a process of increasing automaticity, where deliberate actions are practiced to become faster, less error-prone, and more automatic. It is important to note that explanations of metacognitive skills (and skills in general) are not exhausted by theories of automaticity. Other factors such as cognitive control, flexibility, and metacontrol are also important (Christensen et al., 2016; Pacherie & Mylopoulos, 2021).

However, automaticity does play an important role in skill development, and given the current lack of well-developed competing accounts, focusing on the automatic aspects of metacognitive skill provides a reasonable starting point.

This chapter addresses outstanding questions such as: Does metacognitive skill result from dedicated cognitive mechanisms or from operations that are more domain-general? Can we gain insight into metacognitive skill by examining other successful models of skill? Is there a single parsimonious framework that can explain all categories of skill learning: motor, cognitive, and metacognitive?

### 3.2 Metacognitive Skill

Metacognition is increasingly being referred to as a domain of skill, one that belongs to a larger category that includes both sensorimotor and cognitive skill (Leder & Zawidzki, 2023). Skilled action within any domain entails the high level of control that one possesses over their activity (Mylopoulos & Pacherie, 2021). Metacognitive skill refers to the reliable and flexible capacity to monitor and regulate one's ongoing cognitive activity in the service of one's metacognitive goals (Van der Stel & Veenman, 2010; Lange, 2025a).

Research on metacognitive skill learning, and its neural and computational underpinnings, has largely focused on bottom-up models—where low-level, implicit processes learn by way of stored feedback and reinforcement learning (Proust, 2013; Krueger, Lieder & Griffiths, 2017). While empirical studies have investigated top-down metacognitive learning by way of instructions, such as students being taught to self-monitor and self-regulate their own learning (Zimmerman & Schunk, 2011; Schuster et al., 2020), these studies have largely focused on the effectiveness of various pedagogical strategies. Overall, accounts of explicit metacognitive skill learning have remained largely descriptive and lack a robust explanatory framework for how such skills are acquired and refined.

### 3.2.1 Monitoring and Control

Metacognitive skill presupposes that the components of metacognition, monitoring and control, can improve with practice and training. Metacognitive monitoring refers to the capacity to recognize and identify cognitive states. Monitoring involves the perception of some internal mental states, such as feelings or thoughts, for the purposes of regulating those states or directing behavior. Metacognitive control refers to the active regulation of cognitive states or processes (Flavell, 1979; Wells, 2019). It involves the components of the system that perform mental actions (in contrast to world-oriented actions). Mental actions aim to make cognitive states available that would not otherwise be accessible to awareness or deliberate regulation. The monitoring and control of cognitive activity can involve attention, emotion, planning, reasoning, memory, and various other processes (Slagter et al., 2011; Schraw et al., 2006; Fletcher & Carruthers, 2012; Pearman et al., 2020).

### 3.2.2 Metrics

Metacognitive ability can be quantified using various metrics including neural data, behavioral observation (e.g., task performance), and self-report such as confidence ratings (Fleming & Lau, 2014). Scales for assessing metacognition include the Metacognitive Awareness Inventory (MAI; Schraw & Dennison, 1994), the Metacognition Self-Assessment Scale (MSAS; Pedone et al., 2017), the Metacognition Thinking Skills Scale (Tuncer & Kaysi, 2013), and the Metacognitive Skills Inventory (MSI; Hameed & Cheruvalath, 2021), each targeting somewhat different dimensions of metacognitive capacity.

### 3.2.3 Metacognitive Training

Decades of empirical studies testify to the efficacy of metacognitive training. Research into metacognitive skill learning has a rich history in domains such as reading, mathematics, and general problem solving (Cross & Paris, 1988; Garofalo & Lester, 1985; Davidson & Sternberg, 1998). Metacognitive training has been shown to result in improvements in self-regulation, monitoring, and self-evaluation (McCabe, 2011; Gutierrez de Blume, 2022). Education research indicates that metacognitive skills can be taught effectively, and that doing so is associated with improved student learning and performance across a range of subjects (Girash, 2014). Students with stronger metacognitive abilities tend to solve problems more accurately than students with weaker metacognition (Güner & Erbay, 2021). Metacognitive skill is widely viewed in the education literature as a cornerstone of critical thinking (Kuhn & Dean, 2004).

Metacognitive training plays a significant role in the success rates of Cognitive Behavior Therapy (CBT) and Metacognitive Therapy (MCT). Within both fields, patients are instructed in metacognitive strategies (such as cognitive reframing or mindfulness techniques) for monitoring and regulating their own thoughts and emotions (Dobson, 2013; Fisher, 2021). Research indicates that individuals with metacognitive skills are better able to identify and govern their own harmful thoughts and emotions (Normann & Morina, 2018; Wells, 2019). Conversely, a lack of metacognitive skill can contribute to the preservation of harmful thinking and coping patterns that contribute to anxiety and depression (Cooney et al., 2010; Hagen et al., 2017; Leder & Zawidzki, 2023).

### 3.2.4 Knowledge and Instruction

Metacognitive knowledge, or meta-knowledge, is considered a form of declarative knowledge (Schraw & Moshman, 1995; McCormick, 2003; Stanton, Sebesta, & Dunlosky, 2021). Explicit meta-knowledge takes the form of metarepresentations that are propositionally formatted and refers to some cognitive property or process (Shea et al., 2014; Proust, 2013). Meta-knowledge can simply refer to facts about one's own cognition, such as one's proficiency as a learner, or whether one's attention tends to be easily distracted. Metacognitive knowledge is considered to be distinct from metacognitive skill as it does not automatically lead to the deployment of metacognitive processes (Veenman & Elshout, 1999). Meta-knowledge can also be distinguished from an instruction (Flavell, 1979; Shea et al., 2014). While meta-knowledge refers to facts about oneself as a cognitive agent, metacognitive instructions specify mental actions to be performed. A metacognitive instruction, or meta-instruction, prescribes a mental action directed toward controlling some cognitive process, such as regulating some emotion or focusing one's attention. While decades of research have examined how instruction can guide external behavior in skill development, the process by which metacognitive instruction guides internal mental action has resisted systematic theoretical analysis.

### 3.3 Skill Acquisition

Skill acquisition has been described in psychology and philosophy as a progression from deliberate conscious and declarative rule-following to a nonconscious procedural stage where aspects of performance become more automatic, fast, and accurate (Anderson, 1982, 2020; Dreyfus & Dreyfus, 1986; Fitts & Posner, 1967; Kim & Ritter, 2015). This framework has been used to help explain skill acquisition within both the motor and cognitive domains. Here, we submit that this framework can also be used to understand the acquisition of metacognitive skill. Fitts (1964) proposed that motor skill acquisition unfolds across three phases (Figure 3.1): the cognitive phase, the associative phase, and the autonomous phase. In the *cognitive phase*, learners rely heavily on attention and explicit knowledge to understand task goals and encode an initial action plan, resulting in slow and error-prone performance. In the *associative phase*, performance becomes more consistent as learners use feedback to refine the movement pattern and reduce errors. In the *autonomous phase*, execution becomes largely automatic and requires less conscious control, while performance can continue to improve in efficiency, accuracy, and adaptability.

Anderson (1982) expanded on Fitts' (1964) framework with a three-stage model of cognitive skill acquisition instantiated in the computational cognitive architecture ACT-R (Anderson & Lebiere, 1998). The first stage is referred to as the declarative stage (corresponding to Fitts' cognitive stage), where the learner receives instructions about the skill. The second stage involves knowledge compilation and proceduralization (corresponding to Fitts' associative stage), which entails the gradual process of transitioning from declarative knowledge to procedural knowledge. This second stage is really a transition between the first stage and the later stage. The final stage, referred to as the procedural stage (corresponding to Fitts' autonomous stage), involves further refinement of procedural knowledge and a gradual speedup of performance.

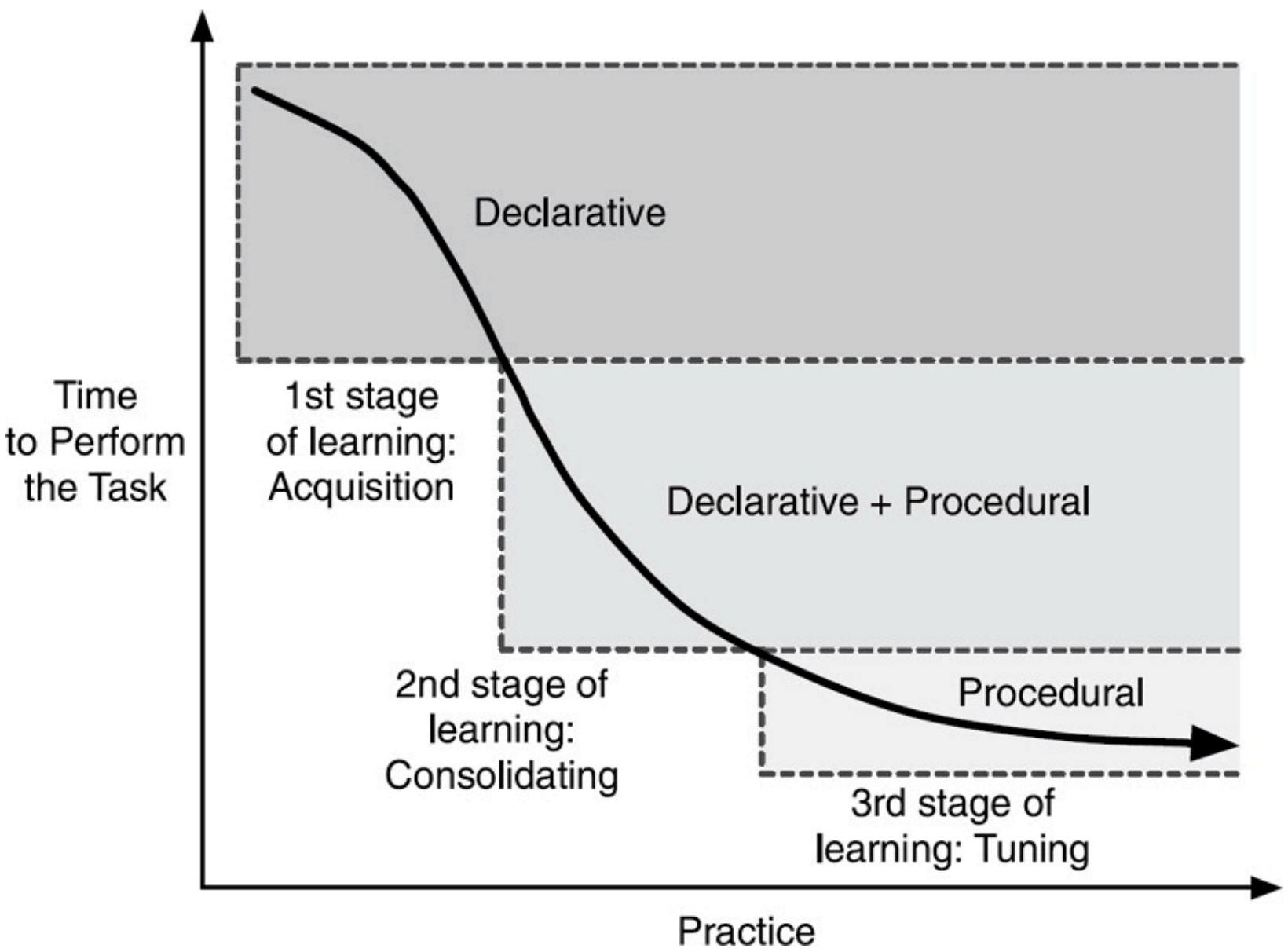

**Figure 3.1.** Performance changes during three stages of skill acquisition (Kim & Ritter, 2015).

### 3.3.1 Production Rules

ACT-R fundamentally distinguishes between procedural and declarative knowledge, which Anderson (1982) relies on for explaining the underlying mechanisms involved in cognitive skill learning. Anderson's framework is consistent with accounts of skill in neuroscience and philosophy, which are likewise grounded in the literature on declarative and procedural memory (Squire, 1992; Christensen, Sutton & McIlwain, 2016). Declarative knowledge is propositionally-formatted and structured within semantic networks. Procedural knowledge is implicit, non-propositional and also referred to by skill researchers as "procedural representations" (Pavese, 2019). In Anderson's model, procedural representations are computationally specified as "production rules" which are a dominant form of representation within accounts of skill (Newell, 1990; Taatgen & Lee, 2003; Anderson et al., 2021). Production rules, also called "productions", transform information and change the state of the system to resolve a problem or complete a task. A production rule is modeled after a computer program instruction in the form of a "condition-action" pairing (Figure 3.2), which is essentially a "pattern-directed invocation of action" (Stocco et al., 2021). It specifies a condition that, when met, performs a prescribed action. A production is also thought of as an "if-then" rule. *If* the condition it specifies is satisfied, *then* it fires an action. Production rules (procedural knowledge) are considered to be central to human intelligence, and fundamental to the realization of cognitive skills (Anderson, 1993).

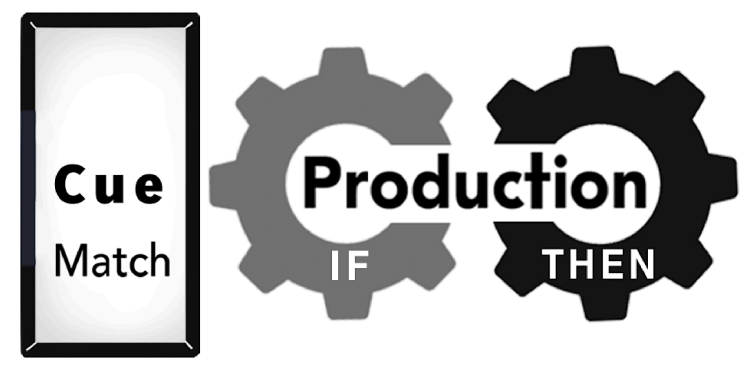

**Figure 3.2.** Production rules are formatted as an *if-then* rule, or condition-action pairing. *If* the condition side matches to the cue in working memory, *then* it fires an action.

This is clarified by noting that specific procedural knowledge (productions) is generally not innate in humans. For example, a child must develop procedures to print their name (motor action), perform mathematical calculations (cognitive action), and regulate their focus (metacognitive action). They must learn that prompts and conditions such as "print name," "solve for x," or "pay attention" are paired with the appropriate action sequences. Once these actions are associated with the correct cues, practice is required to refine the supporting production rules and to improve performance.

With sufficient practice, these productions become stored in procedural memory. When a relevant cue appears in working memory ('print,' 'calculate,' 'focus'), matching productions will activate and execute the correct actions. In this way, cues in working memory can trigger procedural knowledge across motor, cognitive, and metacognitive domains.

### 3.3.2 Proceduralization

The transition from actions guided by explicit declarative knowledge to actions controlled by implicit procedural representations is referred to as proceduralization (Fitts, 1964; Ford, Hodges & Williams, 2005; Kim & Ritter, 2015; Anderson, 1982, 2020). This notion has been useful in explaining the acquisition of physical skills in perceptual-motor expertise (Beilock & Carr, 2001) and cognitive skills such as mathematics (Tenison & Anderson, 2016). A key feature of proceduralization is that, as declarative knowledge is retrieved and repeatedly practiced, the inefficient knowledge retrieval can be skipped. This results largely from procedural knowledge becoming associated with the cue itself, and gradually relying less on the slow retrieval of declarative knowledge. As a result, the performance time speeds up and working memory load decreases. Task performance can be further refined by mechanisms such as time-delayed learning, where faster productions are rewarded.

The building and refining of procedural knowledge marks a significant point of convergence across models of skill learning. This principle has also been extended to metacognition, where researchers have proposed that metacognitive improvement involves refining the procedural knowledge people use to monitor and control their own cognitive processes (Schraw & Moshman, 1995; Kotseruba & Tsotsos, 2020; Wu, Oltramari, & Ritter, 2025).

### 3.3.3 Dual-Process Metacognition

According to dual-process or "dual-system" theories, metacognition broadly comprises two kinds of processes: conceptual metacognition and procedural metacognition (Koriat & Levy-Sadot, 1999; Arango-Muñoz, 2011; Shea et al., 2014; Proust, 2019). Type 2 or "System-2" conceptual metacognition involves the use of explicit metarepresentations with semantic content that guide monitoring and control processes. Type 1 or "System-1" procedural metacognition involves implicit monitoring and control processes that are driven by non-propositional representations such as feelings of knowing (FoK), confidence, and fluency.

These two types of metacognition exhibit characteristics of dual-process theories more generally. Type 2 metacognition is considered to be slow, effortful, knowledge-driven, and requiring working memory. Type 1 metacognition is considered fast, implicit, affect-driven, and automatic. Within this framework, skill acquisition is often modeled as a sequence wherein Type 2 processes "migrate" to become Type 1 operations (Kahneman & Frederick, 2004; Dayan, 2009). Likewise, metacognitive skill learning can be largely understood as a process by which Type 2 metacognitive processes compile or transition to become Type 1 metacognitive operations (Conway-Smith & West, 2022).

### 3.4 Metacognitive Proceduralization

The research discussed so far has provided the background for our proposal that metacognitive skill learning develops largely by way of proceduralization. Here, we submit that metacognitive skill progresses from an early declarative stage of instruction-following to an expert procedural stage of increased automaticity (Figure 3.3). This framework helps account for the signature properties of skill exhibited by metacognition in the later stages of skill learning, as performance becomes faster, more automatic, less reliant on working memory, and operating under reduced levels of conscious access.

#### 3.4.1 Declarative Stage

The metacognitive novice begins with explicit instructions to monitor or control a cognitive state (e.g., attention, emotion, learning). These instructions are typically introduced through verbal or written guidance. Initially, an internal or external stimulus cues the novice to monitor or regulate a cognitive state, which prompts the retrieval of the relevant instructions into working memory. The deliberate execution of metacognitive instructions triggers the activation of procedural knowledge (production rules) that will enact them step by step. Early metacognitive performance therefore exhibits the classic characteristics outlined by Fitts and Anderson in the declarative stage: it is slow, effortful, error-prone, and heavily dependent on working memory resources.

#### 3.4.2 Associative Stage

The metacognitive intermediate has gained modest experience practicing instructions across repeated trials. As a result, they have built a substantial amount of procedural knowledge for carrying out these instructions in context. With continued practice, strong associations form between this procedural knowledge and recurring task conditions (or other task-relevant cues), and performance begins to bypass the slower retrieval of declarative instructions. Because this more direct route to procedural knowledge is faster and more efficient, it is reinforced and becomes increasingly likely to be selected in the future. At this point, metacognitive performance shows many of the features characteristic of the associative stage: monitoring and control become faster and more effective, require less effort, and operate with increasing automaticity.

### 3.4.3 Automatic Stage

The metacognitive expert is able to carry out monitoring and control processes quickly and effectively, often with little perceived effort. The presence of an initial stimulus triggers metacognitive procedural knowledge to be deployed automatically, in a cue-driven way. Procedural knowledge bypasses retrieval from declarative memory almost entirely because metacognitive instructions have become embedded within procedural memory—they have proceduralized. As a result, metacognitive expertise demonstrates many of the hallmark characteristics described by Fitts and Anderson in the final stage of skill learning. Metacognitive task performance has become fast, effective, highly automatic, and requiring minimal working memory.

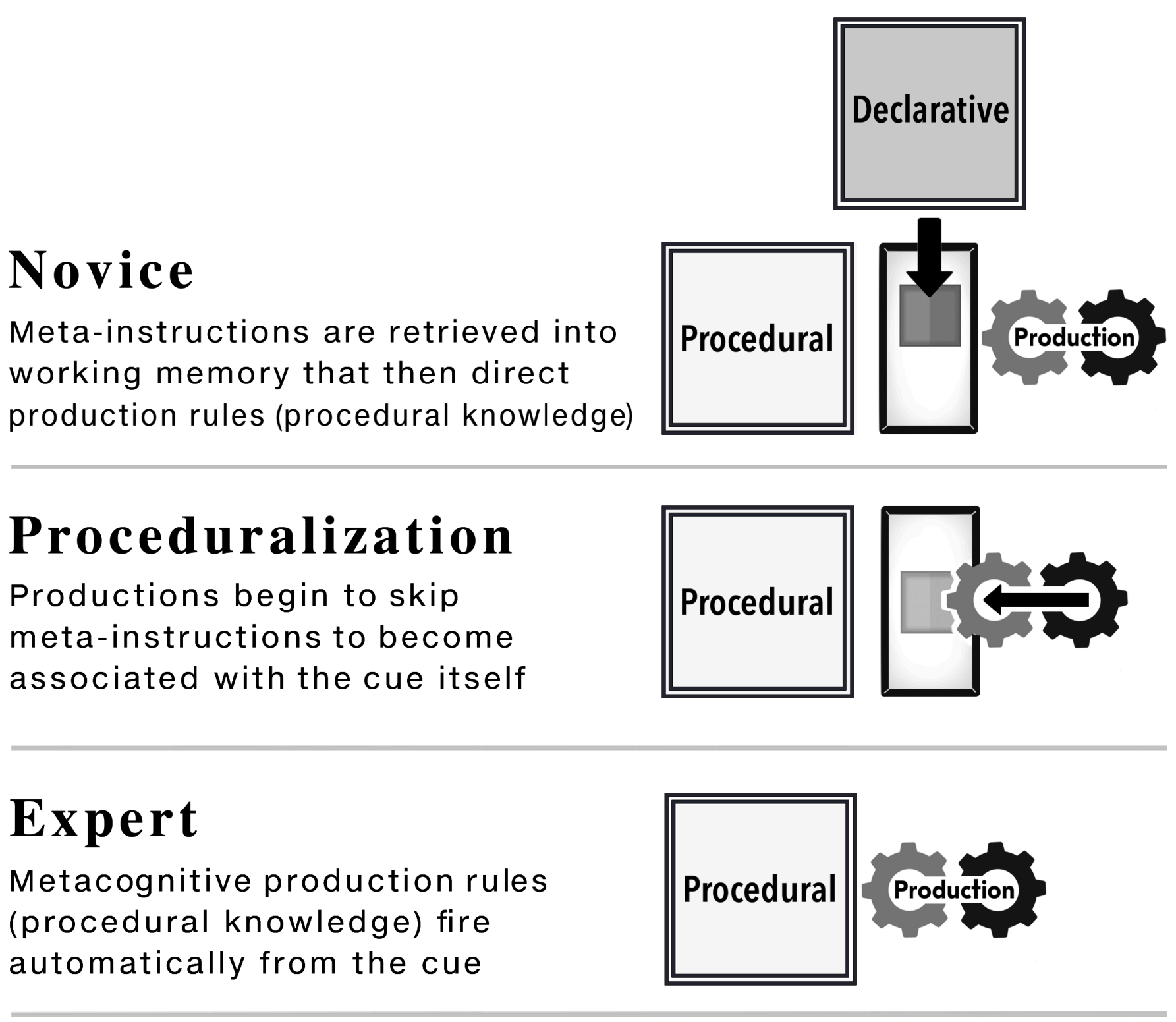

**Figure 3.3.** Three stages of metacognitive skill learning through proceduralization.

### 3.5 Elusive Mechanisms

It is worth considering why proceduralized metacognition has remained largely unidentified. Our reasoning is that it has lain hidden behind two barriers. First, metacognition is largely invisible to outside observers because its actions are less apparent than those of overt motor skill (e.g., tennis) or cognitive skill (e.g., mathematics). Second, automatized skill is generally less perceivable to the performers themselves. As proceduralization progresses, skill mechanisms run largely outside of working memory and operate under reduced levels of conscious access (Beilock & Carr, 2004; Ford et al., 2005).

While declarative knowledge can be directly accessed, procedural knowledge is typically not cognitively accessible, as evidenced by findings that procedural representations are largely unavailable for verbal report (Squire & Zola, 1996; Beilock & Carr, 2001). This is one reason why the development of expert skill often causes performers to report becoming unaware or unable to describe them, as these processes become unconscious habits (Ellis, 1994; Oxford, 2011). This helps to explain why the proceduralization of metacognitive skill has remained elusive. Motor and cognitive skills, when proceduralized, still remain observable. Researchers may watch as a tennis player or math student improves to become faster, more effective, and less error-prone. By contrast, researchers cannot directly observe an individual practicing metacognition, nor can they straightforwardly watch them improve. Moreover, expert performers have less access to their own advanced skills as proceduralization places the underlying processes outside of working memory.

Metacognitive experts may experience a form of what Beilock and Carr (2001) describe in the motor domain as "expertise-induced amnesia." Even though many skill types do not fully automate and continue to require some declarative knowledge, their conscious accessibility is significantly reduced. Therefore, we suggest that metacognitive proceduralization has remained elusive because it has been concealed behind twin blindfolds: that of the researcher and that of the participant.

### 3.6 Proceduralized Emotion Regulation

While proceduralized metacognition has not been identified explicitly, its presence has been tacitly detected within the literature. Evidence for its existence has appeared implicitly, often as confounding data. The term "implicit" here refers to something being apparent without being explicitly defined or explained. For instance, there has been significant research into a phenomenon wherein an athlete's performance becomes impaired under pressure, also known as "choking". In high-pressure situations, an athlete can have their skills disrupted by emotional anxiety and inappropriate attentional monitoring (Baumeister, 1984; Masters, 1992; Beilock & Carr, 2005). Here, a paradox can seemingly emerge. Studies indicate that people who tend to be more self-conscious are less likely to have their performance decrease under pressure. In other words, those who routinely feel *more* self-conscious anxiety are *less* likely to choke. What explains these counterintuitive data? This phenomenon has been observed in repeated laboratory experiments by Baumeister (1984) and has since been supported (Lewis & Linder, 1997; Beilock & Carr, 2001). Mixed results are associated with variables such as skill level and task complexity (Wang et al., 2004).

To explain this paradox, Baumeister proposed that those who often experience more self-conscious anxiety have greater practice at self-regulation. This practice, he argues, is what aids performance under pressure (p. 611). While Baumeister's explanation ends there, it contains the signature of proceduralized metacognition: an improvable self-regulation skill that, with sufficient practice, can operate automatically and outside of working memory. On this view, frequent exposure to self-conscious anxiety functions as repeated training trials in regulating emotion and attention. Through long-term practice of emotion regulation, metacognitive control processes can be learned to the point of automaticity, becoming more efficient and requiring minimal attention (Richards & Gross, 2000; Vohs & Baumeister, 2016). As a result, these control routines can help stabilize emotion and attention during performance, mitigating the effects of pressure and supporting skilled action. Here, we see that what initially appears as confounding data becomes more interpretable when viewed through the lens of proceduralized metacognition.

### 3.7 Proceduralized Attentional Control

Attentional control involves concentrating on an object of focus while ignoring irrelevant stimuli (Lutz et al., 2008). Learning attentional selection (i.e., how targets are prioritized for processing) has been studied from both bottom-up and top-down perspectives (Corbetta & Shulman, 2002; Theeuwes, 2010). Some have argued that this is a "failed dichotomy" and have called for more integrated models (Awh, Belopolsky, & Theeuwes, 2012). A previously unrecognized mechanism of learning has suggested a limitation in prevailing theories.

Recently, researchers have investigated a unique way that attention can be automated to become an attentional "habit" (Anderson, 2016; Salovich, Remington, & Jiang, 2018). These studies suggest a form of learned attention that can direct processing resources with little cognitive supervision. Whether this reflects bottom-up processes, top-down processes, or some combination remains controversial. One way to clarify the issue is to treat attentional control, when directed at internal states, as a metacognitive process that is susceptible to proceduralization.

Experiments by Ramamurthy and Blaser (2017) demonstrate what they call "procedural attention." In their studies, participants performed visual tasks while receiving explicit instructions about where to direct attention. After training, they completed a task designed to prevent both bottom-up cueing and top-down selection. Even in their absence, the trained allocation pattern persisted, with attention continuing to go to the rehearsed locations. The authors take this as evidence for an "offline" mode of attentional selection, one that is cognitively unsupervised and automatic. As they put it, "analogous to the procedural memory that guides skilled motor behavior, one can acquire new selection rules that are flexible and context-dependent, yet also implemented automatically and without supervision—a kind of procedural attention" (p. 1).

The researchers' data indicate that subjects' attentional control was proceduralized in stages which correlate with those described by Fitts (1964) and Anderson (1982). Ramamurthy and Blaser (2017) characterize the process as one of initial rule-following that, with practice, becomes ingrained in procedural memory. Participants begin with declarative instructions for how to allocate attention, rehearse them across trials, and eventually implement them automatically through procedural knowledge. If metacognitive monitoring relies on internally directed attentional processes, then these findings suggest a mechanism by which those processes could be proceduralized through practice.

### 3.8 Flexibility

Many accounts of skill treat flexibility as central to expertise (Fridland, 2017; Pacherie & Mylopoulos, 2021), and a robust theory of metacognitive skill must account for this. A standard way of distinguishing between skill and mere habit is that skills remain under intentional control and can adapt to context, whereas habits tend to be cue-driven and rigid. Douskos (2019) refers to habits as impulsive routines that dispense with attention because the response is pre-determined, where skills are spontaneous in that they use attention to find solutions in novel circumstances. Skilled behavior, in this sense, exhibits a kind of intelligence that is sensitive to an agent's goals and situational features.

Much of what separates novices from experts is the expert's ability to quickly select effective actions in unusual or changing conditions. When applied to metacognitive control, the contrast is between one's habitual reactivity (e.g., Type 1) that simply repeats a pattern of attention or emotion, and skillful control (e.g., Type 2) that can adapt to changes in one's own cognitive-affective state. For instance, research indicates that deliberate practice of attentional techniques can improve one's ability to allocate attention effectively across diverse contexts (Jahn et al., 2023). By contrast, attentional improvements that develop unintentionally are deployed automatically, but remain context-driven and stimulus-dependent. For example, when attention is habitually directed toward entertainment or phone scrolling, focus tends to stay confined to those contexts (Choi et al., 2021). This kind of habitual automaticity is largely involuntary and lacks the deliberate control and adaptability that characterize expertise. On our view, flexibility in metacognitive skill can be partly described as a type of intentionally automatized mental action that can be deployed adaptively to novel and changing situations.

One reason that deliberate practice supports metacognitive flexibility is that rehearsal can automatize the underlying strategies and thereby reduce working-memory demands, freeing cognitive resources for higher-level control (Schneider & Shiffrin, 1977; Masters, 1992). One implication is that these freed resources can be reinvested in planning, error monitoring, and rapid adjustment to changing situations. Because automatized strategies are less demanding of working memory, they are easier to keep "online" while situational monitoring is ongoing (Posner et al., 2015), and they can be deployed quickly without displacing other task-relevant representations. Just as an experienced driver can adjust to changing road conditions, a skilled metacognitive agent can adapt their mental actions to novel distractions and variations in their cognitive and emotional states, selecting among available control strategies as circumstances shift.

### 3.9 Summary and Discussion

We have offered both theoretical and empirical support for our proposal that metacognitive skill learning and its constituent cognitive mechanisms can be largely explained through a model of proceduralization. The motivation for this claim is to address the theoretical gap that has impeded metacognitive research and its application. Our proposal relies on a prominent explanatory framework employed within models of motor skill and cognitive skill, where declarative task-knowledge is converted into procedural knowledge through practice and further refined. Hence, we propose that a single learning mechanism contributes to the acquisition of motor, cognitive, and metacognitive skill. This, in turn, supports the viewpoint that metacognitive skill need not be treated as an exclusive category of cognitive phenomena.

A goal of science is to unify an array of phenomena within a single theory (Newell, 1990). For this purpose, we have brought together several different threads of evidence and explained them by recourse to a single parsimonious framework. We have attempted to address what Vohs and Baumeister (2016) called a yet unanswered question, "where do these nonconscious self-regulation capabilities come from? How do they develop?" In response we have focused on two paradigmatic cases: emotional control and attentional control. Our model helps reveal that these seemingly separate cases of skill learning are in fact two instances produced by the same phenomenon. There are likely many more instances of metacognitive proceduralization that have been overlooked within the literature, with its signature properties waiting to be identified.

This model also helps to resolve a question posed by researchers Charlton & Starkey (2013), "what effect does proceduralization of attention have on performance?" We submit that proceduralized attention (and metacognition more generally) can operate "offline" and largely outside working memory to automatically assist performance within working memory.

A limitation of this theory is that it does not comprise a full account of metacognitive skill as it exists in advanced experts. A more complete account would include the role of implicit learning, cognitive control, and metacontrol. Instead, this chapter highlights a particular unrecognized aspect of the process of top-down metacognitive skill development. A complementary bottom-up account of metacognitive skill acquisition could involve implicit learning, in which procedural knowledge is shaped and reinforced through direct experience.

This model of metacognitive proceduralization is intended to provide testable hypotheses for follow-up work. One avenue for future research would be to test whether the proposed model and its hypothesized stages align with the empirical data that results from metacognitive training. Another route would be to investigate whether the neural correlates of metacognitive skill learning correspond to the same patterns of activation found within similar studies of motor skill and cognitive skill learning. Chapter 7.3 will detail how to test for the characteristic proceduralization speed-up curve (i.e., the power law of practice).

This chapter has aimed to address a question that remains open for future research: How can human cognition learn to interact with its own processes skillfully?

# Chapter 4. Metacognitive Mechanisms of the Attentional Training Technique

This chapter extends the core theoretical framework developed in Chapter 3 by applying the model of metacognitive proceduralization to a specific clinical intervention: the Attentional Training Technique (ATT). In doing so, it helps to clarify the cognitive processes through which structured metacognitive training improves attentional control and emotion regulation. Here, we argue that the accumulation and refinement of procedural knowledge is critical for disrupting persistent cycles of maladaptive thought and affect. More broadly, this chapter offers a comprehensive metacognitive information-processing theory and can help inform the development of psychotherapeutic interventions.

The material in this chapter is adapted from Conway-Smith and West (2025), "Metacognitive Mechanisms of the Attentional Training Technique," published in the *Proceedings of the 23rd International Conference on Cognitive Modeling (ICCM)*.

## 4.1 Introduction

Understanding the cognitive mechanisms underlying psychotherapeutic interventions is crucial for optimizing their efficacy and refining treatment strategies. The Attentional Training Technique (ATT; Wells, 1990, 2019) has been shown to be effective in alleviating symptoms across various psychological disorders (Rochat et al., 2018). However, the computational and cognitive mechanisms that give rise to its effectiveness remain poorly understood. This chapter investigates these mechanisms using the ACT-R cognitive architecture to provide a more precise account of these processes.

In their 2023 fMRI study, Jahn et al. stated that "understanding the 'how' behind the Attentional Training Technique should lead to a better understanding of attentional control and metacognition in general and could eventually manifest in improved or even more specific treatment" (p. 12). To this end, we aim to articulate the mechanisms of attentional training by applying a model of metacognitive skill that has been effectively used to explore related cognitive processes, including metacognitive sensitivity, emotion regulation, and attentional control (Conway-Smith, West, & Mylopoulos, 2023).

Grounded in the principles of proceduralization, this model provides a computational framework for understanding how metacognitive skills are developed and refined over time, offering a mechanistic account of how training enhances metacognitive monitoring and control through repeated practice and task engagement.

This discussion will help address Wells' (2019) call for a "stronger information processing theory" (p. 13) to explain metacognitive control—its components, functions, and the types of metacognitive information involved in the preservation and disengagement of negative processing. A more precise theoretical account of the Attentional Training Technique's subcomponents may not only enhance its existing applications but also facilitate the development of more effective clinical interventions.

We first provide an overview of the Attentional Training Technique and its practical applications. Next, we outline key aspects of metacognition and the metacognitive skill model that entails proceduralization. We then apply this model to the Attentional Training Technique in an attempt to illuminate its constitutive mechanisms. Finally, we discuss how this refined explanation enhances our understanding of the cognitive processes that support emotion regulation and the alleviation of maladaptive psychological symptoms.

## 4.2 Attentional Training Technique

Psychotherapeutic treatments in metacognitive therapy are grounded in the Self-Regulatory Executive Function (S-REF) model, which explains the role of strategic processes and metacognition in psychological disorders (Wells & Matthews, 1996). The S-REF model posits that maladaptive metacognitive beliefs and knowledge can trigger an adverse thought pattern known as the Cognitive Attentional Syndrome (CAS). The Cognitive Attentional Syndrome is a style of negative processing characterized by inappropriate worry, rumination, and threat monitoring. It involves rigid, self-focused attention that amplifies negative emotions, leading to persistent self-preoccupation and distress. CAS is also associated with maladaptive coping strategies such as thought suppression, avoidance behaviors, and substance abuse (Wells, 2009).

The Attentional Training Technique (ATT) is designed to counteract CAS by enhancing metacognitive control and breaking cycles of negative thought (Knowles & Wells, 2018). Research indicates that the Attentional Training Technique is effective in helping individuals disengage from persistent thinking patterns, interrupt self-focused attention, and strengthen metacognitive awareness (Knowles et al., 2016; Nassif & Wells, 2014).

Neuroimaging studies have linked ATT to improvements in attentional abilities and changes in brain function (Jahn et al., 2023). However, researchers emphasize that the underlying cognitive mechanisms of ATT remain poorly understood, highlighting the need for a more explanatory framework. Wells (2019) asserts that "a more detailed modeling of the metacognitive and cognitive architectures supporting self-regulatory processing is needed to advance the field" (p. 5). To this end, we submit that the Attentional Training Technique can be largely explained as a form of proceduralized metacognition.

### 4.3 Stages of Metacognitive Skill Learning

We propose that proceduralization is key to understanding attentional training as a subdomain of metacognitive skill. Metacognitive proceduralization articulates a mechanism by which monitoring and control become more skillful across domains such as attention, emotion, and metacognitive sensitivity (Conway-Smith, West, & Mylopoulos, 2023; Conway-Smith & West, 2024).

Metacognitive proceduralization posits that metacognitive skill develops from an initial stage of instruction-following to an advanced stage where performance largely relies on automatic procedural knowledge (production rules). In the later stages, monitoring and control processes are deployed quickly, more automatically, and require less working memory. This shift towards automatization not only enhances the efficiency of cognitive processes but also frees up cognitive resources, allowing for more flexible metacognitive operations.

According to the theory of metacognitive learning via proceduralization, a metacognitive practitioner progresses through *three stages of training*, which we propose as a structured framework for understanding the Attentional Training Technique.

**Novice Stage.** Training begins with the practitioner being provided with metacognitive instructions that direct attentional control processes toward a specific target of focus (e.g., a visual point, physical area, or sound). These instructions are stored and retrieved from declarative memory and executed by production rules. Performance at this stage is slow, effortful, error-prone, and demands significant working memory resources.

**Intermediate Stage.** Through proceduralization, repeated practice refines instructions into faster, task-specific production rules, reducing reliance on declarative knowledge. These specialized productions for attentional control are rewarded and reinforced, allowing metacognitive performance to become quicker, more automatic, and less cognitively demanding.

**Expert Stage.** Metacognitive instructions for directing attentional control in this context are almost fully converted into procedural knowledge and stored within procedural memory. Upon encountering a relevant stimulus (e.g., a cue to focus or re-focus attention), the appropriate production rules activate automatically, requiring minimal conscious effort. At this stage, attentional control is fast, efficient, and highly automatic, thereby demonstrating several hallmark characteristics of expertise.

### 4.4 Empirical Support for Proceduralized Attention

Empirical findings support the idea that attentional control can be proceduralized through training. For instance, Ramamurthy and Blaser (2017) introduce the notion of "procedural attention" to describe how practiced, instructed patterns of attentional deployment become automatically executed in an "offline" mode, which they state is analogous to the procedural memory that guides skilled motor behavior. Relatedly, recent work on "attention habits" argues that selection history can shape attentional deployment in durable, semi-automatic ways that nonetheless remain susceptible to strategic control (Anderson et al., 2026).

Additional evidence for proceduralization comes from data consistent with the power law of skill acquisition. Logan (1988) operationalized automatization as a speed-up in reaction times (RTs) that follows a power function—characterized by rapid initial improvement followed by a gradual leveling off. This negatively accelerating learning curve has been widely observed in both motor and cognitive skill domains (Newell & Rosenbloom, 1981; Anderson, 1982). Evidence that attentional control improves with practice according to a power law was provided by Shin et al. (2015). In a multi-session rapid serial visual presentation (RSVP) task, participants showed gains in target identification and reductions in attentional blink, both following a negatively accelerating curve typical of procedural skill learning. These findings suggest that attentional control, like other motor and cognitive skills, develops through structured practice and proceduralization.

### 4.5 Clarifying the Attentional Training Technique

Building on the theory of metacognitive learning via proceduralization, we apply this framework to identify the key features and stages of proceduralization in the Attentional Training Technique (ATT).

Jahn et al. (2023) describe the ATT method as implemented via a standardized audio protocol based on the Metacognitive Therapy (MCT) manual (Wells, 2009). Participants receive instructions on directing their attention while listening to six simultaneous audio stimuli: a bell, traffic noise, birds, rushing water, crickets, and a ticking clock. Each 12-minute session consists of three phases: selective attention (focusing on one sound at a time), attentional switching (shifting between sounds), and divided attention (attending to multiple sounds simultaneously).

Participants practiced ATT twice daily for five days. By the study's conclusion, they exhibited improved performance, demonstrating faster, more accurate, and flexible attentional control compared to the control group. This training trajectory aligns with the transition from effortful, declarative instruction-following to more automatic proficiency that would be characteristic of metacognitive proceduralization.

Alternative explanations of attentional training have emphasized mechanisms such as reinforcement learning (Krueger et al., 2017) and predictive coding (Clark, 2015). While these accounts provide valuable perspectives, they do not adequately explain the cognitive mechanisms by which deliberate attentional control transitions to a form of automatic regulation that remains flexible.

### 4.5.1 Exiting Maladaptive Thoughts

Metacognitive proceduralization also suggests a useful framework for understanding how attentional control training flexibly mitigates symptoms of psychological disorders. The Self-Regulatory Executive Function (S-REF) model describes how disorders such as anxiety and depression involve perseverative negative thinking (e.g., worry, rumination), characterized by repetitive cognitive loops (Wells, 1995, 2000). Cognitive Attentional Syndrome (CAS) exemplifies this process, where threat-related thoughts become self-reinforcing without an exit condition from the loop.

The proposed computational perspective helps to clarify how attentional training disrupts maladaptive cognitive cycles. In ACT-R, production rules (procedural knowledge) operate as condition-action pairs. *If* a condition in working memory is met, *then* a prescribed action is executed. At the initial stage of the Attentional Training Technique, practitioners act out declarative instructions to attend to specific stimuli and disengage attention from others. Through repeated practice, production rules act out these instructions to form associations that more quickly recognize maladaptive thought patterns, and more effectively deploy strategies to disengage from them. This is consistent with clinical insights on the value of developing a form of metacognitive discernment (or meta-awareness) that can "identify thoughts as thoughts" (Moore, 1996), which is a crucial step toward breaking repetitive cycles of negative thinking. For example, an individual prone to rumination may, through attentional training, develop metacognitive productions that detect and disengage from intrusive thoughts, effectively providing an exit condition (Figure 4.1).

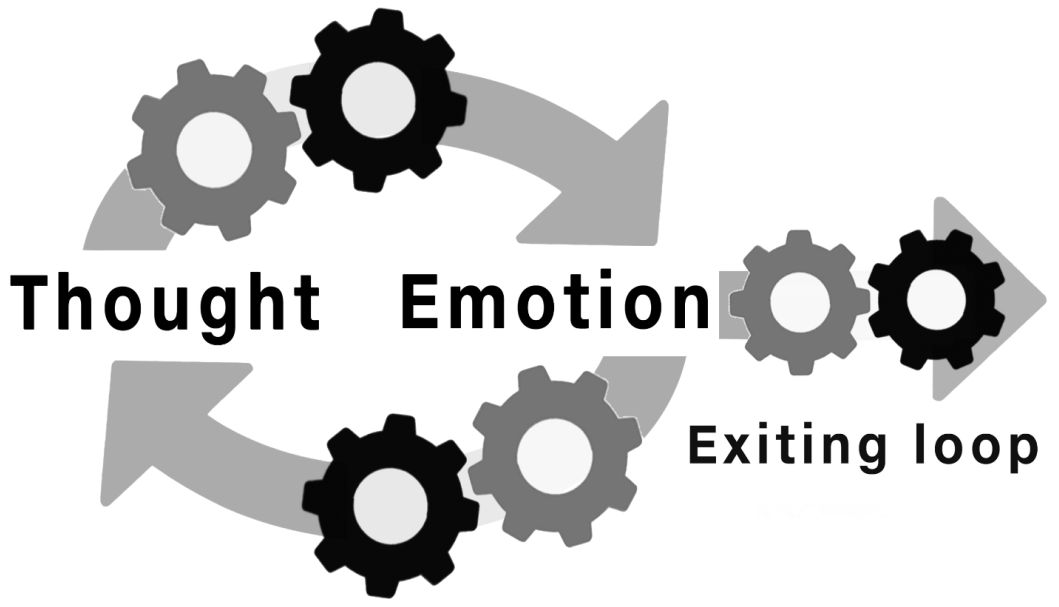

**Figure 4.1.** Maladaptive thought and emotional loops can persist without an exit condition. Attentional training develops production rules that recognize and disengage from these patterns, providing an exit condition from the loop.

### 4.5.2 Exiting Negative Emotions

Attentional training has also been shown to reduce maladaptive patterns of negative emotion, but the underlying mechanisms are not well specified (Wadlinger & Isaacowitz, 2011; Wells, 2019). We propose that the same procedural mechanism that supports exiting maladaptive thought cycles can likewise support exiting maladaptive emotional cycles.

This proposal is consistent with the idea that both declarative knowledge (explicit, propositional states) and affective signals (implicit, non-propositional states) can be represented as patterns of information within working memory, and therefore can be targets for production rules (West & Conway-Smith, 2019). On this view, productions can learn to match not only to conceptual states (e.g., worry) but also to affective patterns (e.g., anxiety), and through practice, trigger attentional shifts that disengage from those states.

This shared mechanism suggests a unified account of how metacognitive training can interrupt loops of both maladaptive thought and emotion. In computational terms, exiting a repetitive loop requires an explicit exit condition, implemented within the ACT-R framework as an exit production. In this case, an exit production fires when an emotional pattern is detected and a control response is initiated (e.g., attentional disengagement, reorienting to task, or another tactic). Over time, training can strengthen the cue–action associations that support these exit productions, increasing their speed, reliability, and context sensitivity, and thereby reducing the likelihood that negative emotional episodes become prolonged or recurring.

### 4.6 Theoretical Support for Transfer Effect

This analysis also helps to explain why attentional training appears to produce transferable gains, with improvements carrying over to other tasks and domains beyond the training context (Ducrocq et al., 2016; Chua et al., 2021). Proceduralization provides a clear mechanism for this transfer by specifying the representational units that change with practice. On this account, attentional training does not merely improve performance on the specific attentional exercise, it builds and refines metacognitive productions that implement general control operations, such as selectively orienting, switching, dividing, and re-stabilizing attention.

The role of production rules in skill transfer has been studied extensively in cognitive skill learning. As procedural knowledge becomes more efficient and more broadly cueable, it can support both near and far transfer across tasks that share functional demands, even when superficial features differ (Singley & Anderson, 1989; Taatgen, 2013). From this perspective, transfer occurs when the trained productions match to abstract task-relevant cues (e.g., “attention has drifted,” “distractor is detected”) rather than to stimulus-specific details (e.g., a specific sound or affect type). Once these rules are compiled and strengthened, they can be triggered quickly and with minimal working-memory load whenever similar control is required.

This account helps to explain why attentional control can become more stable and less effortful across various settings, without requiring explicit instruction for each new task. More broadly, metacognitive skill follows the same principles of generalization. Training broad subdomains of metacognitive control can produce reusable control resources that can redeploy across a range of contexts. In this way, proceduralization offers a principled explanation of how some psychotherapeutic interventions can produce broader cognitive benefits: training strengthens general-purpose metacognitive control strategies, which can be deployed wherever shared demands are present.

### 4.7 Future Directions and Implications

Further empirical work is needed to test and refine this chapter's claim that metacognitive proceduralization underlies the Attentional Training Technique (ATT). First, model validation is critical. Task-based fMRI and EEG studies may also reveal proceduralization-related changes analogous to those observed in motor and cognitive skill learning, including reduced frontal theta power and diminished prefrontal activation. Second, the extent to which ATT produces domain-general versus task-specific improvements remains an open question. Clarifying this distinction will be essential for optimizing its application across clinical, educational, and high-performance contexts. Third, integrating insights from metacognitive reinforcement learning and research in utility learning (Krueger, Lieder, & Griffiths, 2017; Wu et al., 2025) may enhance the model's ability to represent the dynamics of proceduralization over time. Fourth, future work could test the hypothesis that behavioral measures of attentional control—such as visual gaze stability, reaction time, and accuracy—follow a power law of learning, with rapid initial gains tapering off with continued practice.

## 4.8 Conclusion

This chapter has provided a computational account of the metacognitive mechanisms underlying the Attentional Training Technique (ATT), characterizing ATT as a case of metacognitive proceduralization. Framed within the ACT-R cognitive architecture, the account clarifies how repeated practice can transform effortful attentional strategies into automatic procedural routines. On this view, proceduralization helps explain how ATT improves attentional control and emotion regulation by strengthening production rules that support disengagement from maladaptive thought and affective loops. More broadly, because these procedural control routines can be cued across contexts, the framework also offers a principled explanation of why attentional training may generalize beyond the specific training task.

# Chapter 5. Metacognitive Threshold: A Computational Account

This chapter extends the metacognitive proceduralization framework developed in Chapter 3 by applying it to the phenomenon of the metacognitive threshold. Whereas Chapter 4 used the Attentional Training Technique as a case for how metacognitive skills can be proceduralized, this chapter examines how training can change the conditions under which internal states become accessible.

The material in this chapter is adapted from Conway-Smith and West (2023), "Metacognitive Threshold: A Computational Account," published in the *Proceedings of the 21st International Conference on Cognitive Modeling (ICCM).*

## 5.1 Introduction

The ultimate goal of Cognitive Modeling is to build a Unified Cognitive Architecture that can simulate most, if not all, human cognitive abilities (Newell, 1990). Cognitive architectures such as ACT-R and Soar (Anderson & Lebiere, 1998; Laird, 2012) have achieved notable success in modeling knowledge-driven behavior; however, to date there is a scarcity of models related to phenomena surrounding metacognition. Modeling empirical findings regarding metacognition is an important step toward more complete and accurate theories of human cognition that better account for reflective processes.

This chapter addresses a cognitive phenomenon referred to as the metacognitive threshold—the minimum level of stimulus needed for a mental state to be perceived. Specifically, we will address the variability of the metacognitive threshold, which can be reliably lowered to give an agent improved perceptual access to their own internal cognitive states (Pauen & Haynes, 2021). The degree of an individual's introspective acuity is also referred to as "metacognitive sensitivity." This can be reliably improved, and the metacognitive threshold can be lowered by way of metacognitive training such as employing specific mindfulness techniques (Fox et al., 2016). In cognitive psychology, mindfulness is defined as deliberate attention directed toward perceptible mental experiences, such as affective states, sensations, thoughts, etc. (Holas & Jankowski, 2013; Lange, 2025a).

Greater access to and control of one's own mental states have been shown to strongly correlate with improved psychological health and overall cognitive functioning (Grossman et al., 2004; Rigby et al., 2014). While decades of research support the effectiveness of metacognitive techniques in influencing one's metacognitive threshold, the underlying cognitive mechanisms remain poorly understood. At present, there is no clear cognitive or computational account of how the metacognitive threshold is raised or lowered through training, attention, or sustained introspective practice.

This chapter will investigate potential computational mechanisms that may contribute to the lowering of the metacognitive threshold. In particular, we will discuss metacognitive techniques that have been shown to increase metacognitive sensitivity, and explore various frameworks for clarifying their underlying cognitive constituents. For this purpose, we will employ the Common Model of Cognition (CMC), originally the 'Standard Model' (Laird, Lebiere, & Rosenbloom, 2017) that provides a unified framework for investigating the fundamental elements of cognitive and metacognitive phenomena. By using the Common Model and specifically ACT-R in this endeavor, we intend to address unanswered questions regarding the architecture and the nature of production rules.

## 5.2 Metacognitive Monitoring as Mindfulness

Metacognitive monitoring and mindfulness are terms that are often used interchangeably within cognitive psychology (Jankowski & Holas, 2014). Scientific interest in mindfulness practice has become a target of interdisciplinary research and has grown significantly over the past few decades (Tang, 2017; Van Dam et al., 2018). Mindfulness involves deliberately focusing on perceptible experiences (sensory, affective, and thought-related) and cultivating a dispassionate awareness of mental states and processes (Brown & Ryan, 2003; Grossman, 2010; Lange, 2025a). Studies indicate that a technique called *detached mindfulness* is a uniquely effective therapeutic practice for developing adaptive monitoring and control over maladaptive cognitive processes (Wells & Matthews, 1994; Wells, 2005).

Detached mindfulness is characterized by the capacity to notice internal states (thoughts and emotions) without reacting to them, that is, without trying to maintain or suppress them. This is cultivated by attempting to perceive moment-to-moment changes in mental events (such as subtle fluctuations in emotion) and letting them pass without further emotional engagement. Mindfulness psychology contends that a significant degree of emotional distress and pathological symptoms are fueled by misperceiving affective experience as more permanent than it actually is. This illusory perception of affective stability has been explained as the result of a high metacognitive threshold (poor metacognitive sensitivity) that prevents subtle affective fluctuations from being detected. Training in detached mindfulness aims to improve metacognitive sensitivity and strengthen one's direct perception of affective impermanence, often referred to as equanimity. In mindfulness therapies that do not promote equanimity, awareness alone may not be sufficient to improve the psychological well-being of practitioners (Cardaciotto et al., 2008). The increased capacity to perceive the momentary impermanence of affective experience is therefore considered a key mechanism in reducing emotional reactivity and promoting adaptive detachment (Tang et al., 2015).

### 5.2.1 Metacognitive Threshold

A psychophysical threshold is defined as the minimum amount of stimulus needed to evoke a perceptual response in a person (Rouder & Morey, 2009). Psychophysical thresholds and their variability have been researched in domains such as sound, vision, interoception, and various others (Kingdom & Prins, 2016). In metacognition research, psychophysical thresholds have been studied in reference to the minimal level of a stimulus required for a person to be aware of some mental state and make a judgment about it (Charles et al., 2020; Sherman et al., 2018; Pauen & Haynes, 2021). These include confidence ratings as well as the momentary fluctuations of affective experience.

Generally, an individual's metacognitive threshold is variable and can be lowered by way of training attention to better perceive the subtle variations of internal cognitive states—equanimity. The training of equanimity through detached mindfulness and meditation practice has shown to be effective at lowering one's metacognitive threshold and enhancing metacognitive sensitivity.

**Metacognitive sensitivity** is the extent to which one can perceive their own mental processes or states, including thoughts, feelings, and emotions (Fleming & Lau, 2014). Mindfulness training can improve metacognitive sensitivity, allowing one to better notice the nuances of their own thought and affective states. Meditation is an umbrella term for practices that employ deliberate attention and engage neurocognitive processes that can produce advantageous effects on brain and behavior (Baird et al., 2014). Different metacognitive strategies and meditation techniques can target different cognitive outcomes, such as narrow focus, open monitoring, detachment, or decentering. Various meditation techniques have the reported effect of enhancing metacognitive sensitivity, enabling one to perceive a weaker signal strength from internal cognitive states. In developing metacognitive sensitivity, training can improve one's ability to detect subtle variations in emotional stimuli, such as the rapid arising and passing of feelings, thoughts, and emotions.

**Meditation** can involve a variety of practices. Here, we use Vipassana meditation as a useful example. Vipassana meditation (in the tradition of S.N. Goenka) is an old and popular technique that largely focuses on cultivating equanimity: a refined perception and sensitivity to the momentary impermanence of affect and sensations. Regular practice of this technique has shown to result in various cognitive advantages, such as improving executive functioning, enhancing response inhibition, and control over automatic emotional reactions (Chambers, Lo, & Allen, 2008; Andreu et al., 2019). The Vipassana method engages practitioners in guided meditation that directs them to maintain attention on the impermanence of their own sensations (Kakumanu et al., 2018). During this process, practitioners monitor their bodily and affective sensations moment-to-moment, without evaluation, identification, or emotional reactions.

With sufficient practice, practitioners report being able to detect increasingly subtle properties of their own mental states, including variations in affect that were previously inaccessible. In other words, they report greater metacognitive sensitivity and a corresponding lowering of the metacognitive threshold.

### 5.3 Modeling the Phenomena

A computational model that accounts for phenomena surrounding the metacognitive threshold necessarily raises questions about the architecture itself. Which computational components could allow an agent to perceive subtler properties in internal signals? Does this require revising how we think about some of the architecture's basic units? We discuss these questions with specific reference to ACT-R, but the application is intended more generally for the Common Model of Cognition family of architectures. Because our focus is on ACT-R, we refer mainly to production systems. Other CMC architectures, such as Soar, use different, more complex mechanisms, however, the issues raised here remain relevant.

### 5.3.1 Modeling the Metacognitive Threshold

How might production rules account for an enhanced ability to detect internal cognitive signals and their variations? One way is by increasing the speed of production rules. Making productions faster, especially those that notice internal states, increases the likelihood of picking up fleeting or intermittent signals related to emotions and epistemic feelings, such as confidence and feelings of knowing (FoK). A complete model of this phenomenon would need to represent internal signals, how they are detected, how they enter current awareness, and how training improves this process. Here, we focus only on how production-rule acceleration could occur (see West & Conway-Smith, 2019, on how affect and epistemic feelings can be incorporated into this kind of model).

With regard to production rule speed-up, at least four mechanisms could accomplish this:

*1. The ticking clock mechanism*

Production rules fire when a fixed amount of time is up. The use of this mechanism produces production timing that is analogous to the intervals of a ticking clock. The timing of this process is based on neural functions that are generally considered to occur within the basal ganglia. The timing for production rule firing is generally estimated to be 50ms (Stocco, 2018). During this interval, productions that match the buffer conditions are identified. When this time is up, the matching production with the highest utility will fire. Using this mechanism, production time could be sped up by shortening the clock speed. This could occur through top-down feedback related to attention, as its influence has been observed in other psychophysical thresholds, such as improving perceptual sensitivity (MacLean et al., 2010).

*2. The fire when ready mechanism*

Production rules fire when they are ready. ACT-R is essentially a fire-when-ready model. ACT-R assumes that it takes 50ms for a production to fire, but if no production rule matches the buffer conditions, ACT-R will wait until the buffer conditions change. For example, for memory retrieval, ACT-R waits for the knowledge chunk to be delivered into the declarative memory buffer and then fires the matching production. Hence, the overall time taken is the memory retrieval time plus 50ms. However, if an alternative production matches the buffer conditions before this occurs, then it will fire instead. Using this mechanism, production firing can be made faster by using productions that do not wait for information from memory or perception. These types of productions can be generated through the production compilation mechanism in ACT-R.

3. *The narrow focus mechanism*

ACT-R is capable of multitasking and even mind wandering if the appropriate productions are available. The simplest way of producing a faster rate of firing for a specific type of production is to maintain the buffer conditions such that only this type of production can fire. Under these conditions, ACT-R can be said to model a narrow focus of attention.

4. *The faster production mechanism*

Some productions may be faster in their capacity to match and fire than others. Productions range in the complexity of their internal actions (Taatgen, 2013). The consequences of this at the neural level could imply that more complex productions take longer than simpler productions.

Stewart et al.'s (2010) neural model of the basal ganglia estimates that this would produce a range between approximately 34ms-44ms for simple productions, and 59-73ms for complex productions. If this is the case, then simpler productions would speed up the firing time.

Each of these mechanisms could lower the metacognitive threshold by speeding up productions and thus increasing fidelity. To be clear, this viewpoint does not require that a threshold exists *in fact*, only that the resulting effect would appear to be so. Hence, from an architectural standpoint, this particular issue is not regarding thresholds but rather the complexity and timing of production rules. We propose that increasing the rate of production rule firing can potentially account for reports of increased metacognitive sensitivity as a result of metacognitive training, and that Common Model type architectures can model this.

## 5.4 Metacognitive Proceduralization

The process by which simpler, faster production rules are developed through metacognitive training can be largely explained by way of metacognitive proceduralization (Conway-Smith, West, & Mylopoulos, 2023). Proceduralization is a concept used in the skill acquisition literature to explain the cognitive mechanisms involved. It refers to the process by which a task or skill becomes automated, allowing it to be performed more efficiently and accurately, with minimal perceived effort and attention. The process involves converting slow declarative knowledge into fast procedural knowledge that is increasingly refined. Performance can be further improved by mechanisms such as time-delayed learning, where faster productions are rewarded. Within this framework, metacognitive skill develops through three stages, similar to those of Fitts and Posner (1967) and Anderson (1982), from an early stage of instruction following to an expert stage that relies on refined, automatic procedural knowledge (production rules).

The practitioner of detached mindfulness can be understood as progressing through three stages of training, through which metacognitive sensitivity improves and the metacognitive threshold is lowered:

**The novice stage** begins with metacognitive instructions that direct monitoring and control resources in a specific way. In the case of metacognitive training in equanimity, these instructions direct the novice's attention toward momentary fluctuations in affective experience (i.e., a feeling, sensation, or emotion). The instructions are carried out by production rules that retrieve them from declarative memory and execute them. Here, production speed-up could occur through mechanism 3 and possibly mechanism 1.

**The intermediate** stage of metacognitive training involves proceduralization, where instruction-following results in the creation of faster production rules for accomplishing the task. Specifically, repeated practice can lead to the compilation of task-specific productions that increasingly bypass declarative knowledge. Because these productions are faster (by avoiding declarative retrieval and, in some cases, by being less complex), they are more strongly rewarded and more likely to bypass instruction retrieval in future instances. Here, speed-up occurs through mechanism 2 and possibly mechanism 4 (with mechanisms 3 and 1 still in play).

**The expert stage** involves a robust accumulation of metacognitive production rules that have been refined and stored in procedural memory. These productions can be deployed automatically to carry out monitoring and control processes quickly and effectively. Here, it is possible that productions accelerated through mechanisms 2 and 4 become so deeply ingrained that metacognitive monitoring and control can occur spontaneously. This would yield an increased ability to monitor even without relying as heavily on mechanism 3 or 1 (though mechanisms 3 and 1 could still increase effectiveness if engaged).

### 5.5 Discussion

This chapter examined how metacognitive training can lower an individual's metacognitive threshold, thereby increasing access to internal cognitive states. We proposed that this process can be explained through metacognitive proceduralization: repeated training improves the efficiency, speed, and reliability of productions involved in monitoring internal signals. On this view, experienced practitioners do not necessarily perceive stronger internal signals; rather, they sample, encode, and act on those signals more efficiently.

The account remains preliminary, and further empirical and computational work is needed to specify which mechanisms are most responsible in different forms of training. Future models should clarify how internal affective and epistemic signals are represented, how they enter awareness, and how changes in production timing interact with attention, motivation, and task demands.

# Chapter 6. Computational Mechanisms of Detached Mindfulness

This chapter develops a computational account of how detached mindfulness can disengage maladaptive patterns of emotional reactivity. Detached mindfulness is a therapeutic technique aimed at observing thoughts and emotions without reacting to them, and it is often described as reducing distress by changing how affect is experienced in the moment. Building on the theoretical framework developed in earlier chapters, this chapter treats detached mindfulness as a form of metacognitive skill that can be trained and refined through practice. It integrates this dissertation's skill-learning account of metacognition with the idea that training can alter the conditions under which internal states become consciously accessible. The goal is to situate detached mindfulness within an information-processing framework, clarifying how metacognitive training can produce more adaptive emotional responses and help explain its broader therapeutic value.

The material in this chapter is adapted from Conway-Smith and West (2024), "The Computational Mechanisms of Detached Mindfulness," published in the *Proceedings of the 22nd International Conference on Cognitive Modeling (ICCM).*

## 6.1 Introduction

The attempt to build a Unified Cognitive Architecture (Newell, 1990) that can replicate human-like intelligence must necessarily account for the routine interplay between affect and metacognitive processes. Historically, cognitive modeling research has focused predominantly on knowledge-based processing such as reasoning, vision, and AI problem-solving, with little or no computational account of the critical role of emotion and metacognition. This need for increased computational understanding is underscored by the fact that perseverative patterns of negative emotion, such as depression and anxiety, are the largest causes of cognitive disability worldwide (World Health Organization, 2022). Consequently, there has been a global push to develop metacognitive techniques that allow individuals to engage with their emotions adaptively.

A particularly effective metacognitive technique is referred to as detached mindfulness (Wells, 2005; Bolzenkötter et al., 2024). This technique focuses on developing one's perception of momentary changes in affective states, which has been shown to significantly reduce distress and emotional reactivity, and improve overall cognitive functioning (Hammersmark et al., 2024).

While decades of clinical research support the effectiveness of detached mindfulness, its underlying cognitive and computational mechanisms remain largely unexplained. This chapter will investigate the cognitive and computational constituents that underpin detached mindfulness and its therapeutic benefits. Specifically, we will discuss the metacognitive mechanisms by which the perception of affective fluctuations deactivates emotional reactivity. For this purpose, we will employ the Common Model of Cognition (CMC), originally the 'Standard Model' (Laird, Lebiere, & Rosenbloom, 2017), which provides a unified framework for investigating the fundamental elements of cognitive and metacognitive phenomena. By utilizing the Common Model, and specifically ACT-R (Anderson & Lebiere, 1998) in this investigation, we intend to address important questions largely unexplored in cognitive models: How does metacognitive training in detached mindfulness reduce persevering patterns of negative emotions? By what computational mechanism does perceiving the momentary changes in affect disengage emotional reactivity such as meta-emotions?

First, we will overview the relevant literature on metacognition and mindfulness techniques. Second, we outline the computational mechanisms involved in a model of metacognitive skill learning. Third, we apply this model of metacognitive skill learning to detached mindfulness to clarify its underlying components and the precise mechanism by which it reduces emotional reactivity as reported in the literature.

### 6.1.1 Meta-Emotion

Meta-emotions are emotions that automatically react to other emotions (Jäger & Banninger-Huber, 2015; Predatu et al., 2020). For instance, a primary negative emotion (sadness) can cause a greater secondary negative emotion (despair) which may cause an even greater tertiary negative emotion (depression). Meta-emotions are instances of positive feedback, in which an emotional response to a primary emotion intensifies the overall emotional experience, leading to an amplified response. Meta-emotions occur as low-level reactive processes that are largely unconscious and involuntary, making them difficult to intervene in.

While therapeutic practices aim to control the resulting effects of meta-emotions such as anxiety and depression, techniques such as detached mindfulness and Vipassana meditation aim to address the source, which is considered the false perception of affective permanence. To date, we lack a mechanistic understanding of precisely how detached mindfulness breaks through the illusion of affective permanence and disengages emotional reactivity. To clarify this mechanism, we will apply a model of metacognitive skill that articulates the components involved in this process and how they interact. Central to this explanation is a process referred to as proceduralization, a framework common among skill theories. We will first discuss the relevant components of metacognition and their expression in the cognitive architecture ACT-R. We will then explore how the components of proceduralization function to produce the therapeutic mechanism active in detached mindfulness.

### 6.2 Metacognitive Proceduralization

Metacognitive proceduralization involves a mechanism by which human cognition becomes more skillful at monitoring and controlling its own processes, such as attention, emotion, and metacognitive sensitivity (Conway-Smith, West, & Mylopoulos, 2023). Previous research has presented proceduralization as a mechanism that can lower the metacognitive threshold, allowing one to perceive increasingly weaker signals from mental states and more subtle changes in affect (Conway-Smith & West, 2023). It is hypothesized that proceduralization accomplishes this through the building and refining of simpler, faster production rules. Faster and less complex productions, particularly those that notice internal states, increase the chances of picking up fleeting or intermittent signals related to emotions and epistemic feelings, such as feelings of knowing (FoK) and confidence. However, this model has not thoroughly addressed the process by which it mitigates emotional reactivity. By extending this research on metacognitive proceduralization, we can investigate a mechanism whereby sufficient metacognitive sensitivity can be developed to deactivate meta-emotions.

As discussed in Chapter 5, metacognitive training in detached mindfulness and Vipassana meditation progresses from explicit instruction-following to more effective, increasingly automatic performance. Across these stages of learning, monitoring and control processes become faster, less effortful, and less dependent on working memory. This, in turn, enables a lower metacognitive threshold, which improves perception of momentary affective changes. This refined perceptual awareness is associated with reduced emotional reactivity, which is central to many contemporary psychotherapeutic techniques (Wells, 2019; Lange, 2025b). The following section specifies a computational mechanism by which training can lower the metacognitive threshold and reduce the presence of meta-emotions.

### 6.3 Deactivating Meta-Emotions

Proceduralization, the development of task-specific production rules, assists in providing a computational account of how training to perceive affective fluctuations (equanimity) results in the deactivation of meta-emotions.

Recall that production rules match and fire off the content of working memory at a default rate of 50ms. That is, productions require at least 50ms to detect a pattern held within working memory. Should a pattern be perceived as sufficiently stable for over 50ms, productions will automatically match and fire off that pattern. Hence, the timing of production rules may be considered a condition of the metacognitive threshold (and psychophysical thresholds more generally) as it provides a partial account of which properties of the stimulus are needed to evoke a response (i.e., strength of signal and perceived stability).

An analogous psychophysical threshold is well known in vision research, where a light that flickers rapidly enough can appear to be constant (Landis, 1954). This visual illusion is exploited in film production, where images presented at 24 frames per second give the sequence the appearance of continuity. The visual threshold at which rapidly changing images appear as seamless motion has been referred to as the "moment of fusion". This visual threshold can be partially raised or lowered due to individual differences such as fatigue and age. For our purposes, the visual "flicker-fusion" illusion is comparable to the illusion of affective stability, in that they both rely on the inability to perceive change above a certain rate.

Similar to the visual threshold, an individual's metacognitive threshold is variable and can be lowered through attention training to perceive weaker signals from internal cognitive states, such as subtle changes in affect. Proceduralization offers a mechanism for developing and refining production rules that are more sensitive to internal signals, so as to eventually break the illusion of affective consistency. A key insight into precisely how the refined perception of affective change deactivates emotional reactivity may come from the timing of production rules.

### 6.3.1 Above the Threshold

To the extent that a person's metacognitive threshold is above the 50ms firing rate of production rules, they will perceive any pattern within working memory to be relatively stable. Should a negative emotion appear to be consistent over the 50ms threshold, productions have sufficient time to match and fire a secondary negative emotion in response to the first. Assuming the same conditions, the secondary negative emotion may be perceived and reacted to again, producing a tertiary negative emotion. As long as this metacognitive threshold remains (along with the illusion of affective consistency) production rules may fire automatically, and emotional reactivity may repeat indefinitely.

This explanation sheds light on a potential mechanism that may generate the continuous increase in negative emotions as experienced in many psychological disorders. Increasing and persistent cycles of maladaptive emotions are among the most common symptoms of mental illnesses and are associated with Cognitive Attentional Syndrome (CAS; Wells, 2009). A nearly universal phenomenon in cognitive disorders, CAS is a style of negative processing marked by fixed, negatively-biased attention which causes maladaptive emotions to be preserved and heightened, resulting in a nearly continuous state of emotional distress.

While there is a lack of computational explanations for the mechanisms underlying this style of maladaptive processing, the timing of production rules can help explain how negatively valenced emotions can be heightened through a process of positive feedback. Production rules also help explain the largely unconscious and involuntary nature of emotional reactions, underscoring the need for metacognitive training to develop productions that counteract them.

### 6.3.2 Below the Threshold

We propose that a key mechanism contributing to the deactivation of meta-emotions is the ability to perceive affective change below the 50ms firing rate of production rules. Reducing the metacognitive threshold below 50ms produces an effect similar to when film speed is reduced below 24 frames per second. The illusion of consistency is broken, and one perceives the rapid arising and passing of momentary experience.

This refined perception of affective fluctuations may inhibit production rules from matching to the rapid changes in working memory (Figure 6.1). In effect, production rules simply do not have enough time to match to buffer conditions. Consequently, as long as sufficient metacognitive sensitivity remains, productions are unable to fire secondary emotions.

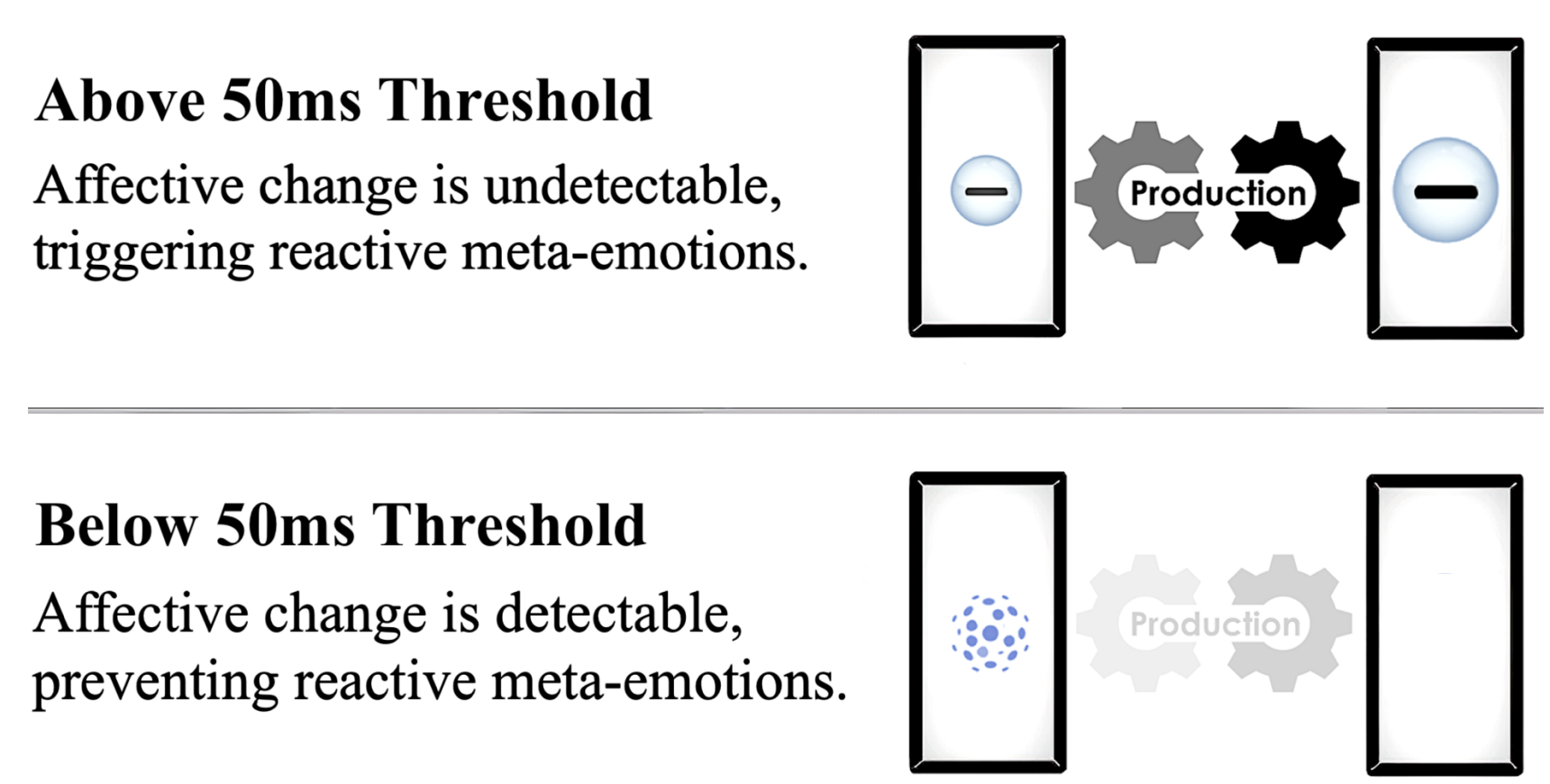

**Figure 6.1.** Threshold-dependent reactivity. *Above the 50ms threshold:* negative emotion in working memory is perceived as sufficiently stable for productions to match and fire a secondary emotion. *Below the 50ms threshold:* perception of emotional change prevents productions from matching and firing secondary emotions.

This lowering of the metacognitive threshold below the 50ms rate requires expert level metacognitive skill, as it necessitates the accumulation of sufficiently refined production rules. These expert production rules are better able to detect subtle variations in affective experience and fleeting signals from other internal cognitive states. Conversely, if one's metacognitive threshold rises above 50ms, the affective pattern may appear stable enough for emotional reactivity to resume.

This account helps articulate how the subcomponents of mindfulness training assist in diminishing cycles of negative emotion in psychological disorders such as the Cognitive Attentional Syndrome. Individuals who suffer from psychological disorders are often caught in patterns of negative emotion without an exit condition from the informational loop (see Figure 4.1). From a computational standpoint, the development of production rules of the type discussed here would provide an exit condition from maladaptive emotional cycles, or prevent the production of negative emotions that would otherwise persist.

This analysis highlights the pivotal role of metacognitive training in emotion regulation and proposes a mechanism by which metacognitive practices such as detached mindfulness may enhance the ability to perceive negative emotions without reacting to them.

### 6.4 Other Considerations

Accounting for mindfulness with cognitive modeling is a multifaceted endeavour, and there are many other considerations. For example, there is the issue of buffer decay, or how long patterns of activity can remain within working memory. These issues would apply to representations of both thoughts and emotion. Another issue is the ability for productions to match to emotional states and to declaratively label them. A particular issue that arises here can be understood in terms of partial matching, or the fidelity of the match. If we take emotion to be a representation of neural activity, then we would expect it to have gradations of variability. Because the ability to recognize emotions would depend on our ability to match to these representational gradients, we would need to assume some form of fuzzy matching. This raises the possibility that some individuals could have more finely tuned productions and conceptual categories for matching emotions, while others may have broader, more fuzzy categories. Chapter 5 argued that increases in production-rule speed could improve sensitivity to the presence of, and shifts in, emotion, and discussed various ways this speed-up could be modeled.

### 6.5 Conclusion

We argued that Common Model cognitive architectures, implemented in ACT-R, can capture key aspects of mindfulness training by modeling it as a form of metacognitive proceduralization that reduces emotional reactivity. While a complete, testable model remains to be developed, a promising direction is to derive neural-level predictions and compare them with neuroimaging findings in trained meditators. Overall, this work helps bridge cognitive modeling and psychological practice by offering a computational account of metacognitive monitoring and control in equanimity training.

# Chapter 7. Discussion: AI and Future Directions

## 7.1 General Discussion

This dissertation has argued that metacognition is best understood as a domain of skill and that improvements in metacognitive control can be explained by recourse to the general mechanisms of skill acquisition. Chapter 2 motivated this claim conceptually and clarified the key features of metacognitive skill. Chapter 3 developed a formal account of metacognitive proceduralization within the Fitts-Anderson framework and ACT-R. Chapters 4–6 applied this account to attentional training, emotion regulation, the metacognitive threshold, and detached mindfulness. Together, these chapters suggest that a variety of metacognitive phenomena can be accounted for under a unified framework in which practice transforms explicit strategies for monitoring and control into increasingly efficient, automatic operations.

This dissertation has clear limitations that are important to acknowledge, several of which have been noted throughout the individual chapters. For instance, this account does not yet sufficiently articulate the implicit learning that can shape metacognitive control through Type 1 processes, such as metacognitive reinforcement learning (Krueger, Lieder, & Griffiths, 2017) or utility-based learning (Wu et al., 2025). However, this lies largely outside the scope of a top-down skill acquisition account (see Section 2.3.4). The proposed framework also emphasizes individual, task-directed metacognition and does not address the social and interpersonal dimensions of metacognitive regulation. In addition, much of the support offered here is conceptual and computational, drawing on empirical findings from previous studies rather than new empirical research of its own. The account is therefore best understood as a structured set of hypotheses about the development of metacognitive skill, intended to guide theory refinement and testing, including the empirical directions outlined in Section 7.3.

## 7.2 Application to Artificial Intelligence

Improving metacognitive capabilities in artificial intelligence is now considered to be crucial to building more efficient, autonomous, and safe AI (Wei et al., 2024; Walker et al., 2025). Despite recent progress, however, engineering robust metacognitive capabilities remains an open challenge (Johnson et al., 2026). The theory of metacognitive skill learning developed in this dissertation offers a framework for informing the design of metacognition in artificial systems, with a central implication that metacognitive control should be learnable and proceduralizable. This theoretical approach has been presented at major AI conferences and workshops such as AAAI (Conway-Smith & West, 2022, 2024) and cited in *npj Artificial Intelligence* (Bergamaschi Ganapini et al., 2025), demonstrating its relevance to contemporary computational cognitive science.

One virtue of grounding this theory in a cognitive architecture is that it specifies computational mechanisms of human self-monitoring and control that can be plausibly implemented in functionally similar ways in artificial systems. This account engages with the history of research in this area by suggesting that metacognition need not rely on dedicated modules or recursion, but can instead be treated as an emergent process arising from the interaction of cognitive modules and information types. Within this framework, the dichotomy between Type 1 metacognition (fast/intuitive) and Type 2 metacognition (slow/deliberative) is replaced by a model in which metacognitive performance is a product of the interaction between working memory, declarative knowledge, and procedural knowledge. From this perspective, high-level reasoning, self-reflective control, and automatic processes form a spectrum of control properties that are expressed by the same underlying architecture. In this sense, “fast” and “slow” processes can be understood as different operating modes of the same architecture, rather than two separate systems.

A key contribution of this theory of metacognitive proceduralization is to treat metacognitive control as a learnable process that current artificial systems do not exhibit in a robust way. At present, AI systems such as LLMs are unable to autonomously refine their own metacognitive strategies (e.g., confidence calibration, error detection, and adaptive strategy switching) over time without humans in the loop (Johnson et al., 2026). Artificial systems cannot reliably determine which metacognitive strategies are best in a context-sensitive way, or when to apply a metacognitive strategy across changing task conditions. Current systems often show weak calibration and error detection, which limits reliable self-correction and autonomous strategy switching. By contrast, human metacognitive learning supports at least two abilities that are relevant for AI design: discerning which metacognitive strategies are best in a particular context, and automatizing effective strategies through practice. Metacognitive learning and automatization allow strategies that are initially computationally heavy, through feedback and repeated use, to become compiled into efficient, context-sensitive control processes, while reserving deliberation for difficult or novel situations.

The present framework offers design principles that go beyond adding a metacognition module or prompting a system to reflect or deliberate further. Metacognitive control should be organized so that computationally inexpensive monitoring signals routinely guide processing, while computationally costly checks are initiated only when necessary. In practice, internal cues such as low confidence, conflict among candidate outputs, or weak evidential support should automatically trigger verification or strategy switching, and should reserve further self-evaluation for cases where uncertainty or task demands require additional computation.

### 7.3 Future Directions

This dissertation developed a theoretical framework for how metacognitive control is refined through practice, and the next step is to obtain objective evidence that specific metacognitive phenomena can be trained and proceduralized. In the skill literature, learning is often reflected in power-law improvement curves (Logan, 1988; Kim & Ritter, 2015). These curves show rapid early gains followed by slower progress as performance approaches an asymptote (see Figure 3.1). Similar patterns have been observed in metacognitive subdomains such as attentional skill and metamemory (Shin et al., 2015; Racsmány et al., 2018), yet metacognition as a whole has not been explicitly addressed. Systematically testing for whether metacognitive training produces power-law learning curves, faster execution, and reduced subjective effort would provide direct empirical evaluation of the present framework. The variety of metacognitive phenomena that could plausibly be studied includes attentional control, emotion regulation (e.g., via detachment and reappraisal), and metamemory. Concretely, future studies can address questions such as:

- Does repeated practice of a metacognitive process produce measurable gains in efficiency (faster implementation, lower perceived effort) and effectiveness (e.g., larger changes in the targeted outcome), consistent with proceduralization?
- Do these gains persist over time, indicating durable skill learning rather than temporary strategy use?

To answer these questions, one promising methodological approach is a multi-session online training paradigm. Participants would be randomly assigned to a metacognition-training group (receiving structured practice of a specific metacognitive process) or to a control group (e.g., read-only or minimal-instruction). A sample of roughly 120 adults would allow for adequate power, subgroup analyses, and robustness checks. The study could follow a short-term longitudinal schedule with one session every two days over two weeks (seven sessions in total), which would balance feasibility with the need to observe learning over repeated practice.

Across sessions, participants in the training group would repeatedly apply a metacognitive strategy to standardized stimuli or tasks, while the control group would be exposed to the same materials without metacognitive practice. Each trial would include: (i) an initial rating of the relevant response (for example, emotional intensity, task confidence, or perceived difficulty), (ii) an opportunity to engage in the metacognitive process, and (iii) post-trial ratings indexing effectiveness (change from baseline) and perceived effort. The reaction time to implement the metacognitive strategy would serve as an index of increasing automaticity, while self-reported perceived effort would provide a measure of cognitive load.

The primary outcomes would be reaction time and perceived effort, while secondary outcomes could include changes in the targeted state (e.g., pre–post test) and individual-difference measures that moderate learning trajectories. A follow-up session would assess retention and the durability of gains. To test for indications of skill learning, power-law functions would be fit to reaction-time data across sessions (for example, using log–log linearization or nonlinear mixed-effects models) and compared against alternative forms such as linear or exponential trends. Similar decreases in reported effort would offer further evidence of reduced cognitive load.

By combining effectiveness ratings (change from baseline), reaction-time measures, and perceived effort within a training-versus-control design, this work would provide evidence that metacognitive training both improves outcomes and becomes more efficient with practice, as expected under proceduralization. An empirical study of this kind would not only evaluate the core predictions of this dissertation but also help inform evidence-based programs that strengthen metacognitive skill for learning, mental health, and everyday self-regulation.

## 7.4 Concluding Remarks

This dissertation has argued that many forms of metacognitive control are best understood as learnable skills governed by general principles of skill acquisition, especially proceduralization. Its central contribution has been to develop a formal cognitive and computational account of how expert metacognitive abilities are acquired. On this view, explicit and effortful strategies for monitoring and regulating cognition can be transformed through practice into faster, more efficient, and automatic forms of control, while remaining capable of deliberate guidance when circumstances are novel or challenging.

By framing metacognition as a domain of skill and situating it within an established computational cognitive architecture, this dissertation provides a unified account of metacognitive phenomena that have been previously investigated separately. Attentional training, emotion regulation, increased metacognitive sensitivity, and detached mindfulness are each clarified here as different expressions of a common learning process.

This account also generates concrete empirical predictions. If metacognitive skill develops through proceduralization, then training should produce identifiable signatures of skill learning, including improvements in effectiveness, reductions in subjective effort, faster implementation, and learning trajectories that approximate the power law of practice. Hence, this dissertation offers more than a mere reinterpretation or clarification of metacognition. It provides a mechanistic framework that can be evaluated, refined, and potentially falsified.

The broader significance of this framework extends beyond research and theory development. It supports applications in education and psychotherapy by providing a more precise understanding of how metacognitive interventions work and why some forms or blends of training produce lasting change. Within cognitive science, it helps connect metacognition with general theories of learning, motor and cognitive control, and skilled performance. In artificial intelligence, it offers a basis for thinking about how capacities for self-monitoring and self-regulation might be designed at scale rather than piecemeal.

This dissertation advances the claim that metacognition and metacognitive skill learning are not mysterious phenomena, but intelligible targets of scientific research and theory. By explaining how metacognitive abilities can be acquired and refined across domains, it lays the groundwork for a more unified science of how minds learn to regulate themselves. Ultimately, I have sought to clarify how intelligence, whether biological or artificial, can more fully realize its potential by learning to direct its own processes skillfully.

**References**

Ackerman, R., & Thompson, V. A. (2017). Meta-reasoning: Monitoring and control of thinking and reasoning. *Trends in Cognitive Sciences*, *21*(8), 607–617.

Alter, A. L., & Oppenheimer, D. M. (2009). Uniting the tribes of fluency to form a metacognitive nation. *Personality and Social Psychology Review, 13*(3), 219–235.

Anderson, B. A. (2016). The attention habit: How reward learning shapes attentional selection. *Annals of the New York Academy of Sciences*, *1369*(1), 24–39.

Anderson, B. A., Lee, D., Yan, N., McKinney, M., & Clement, A. (2026). The Attention Habit II: How selection history shapes the strategic control of attention. *Psychonomic Bulletin & Review.*

Anderson, J. R. (1982). Acquisition of cognitive skill. *Psychological Review*, *89*(4), 369–406.

Anderson, J. R. (1993). Knowledge representation. In *Rules of the mind* (pp. 17–44). Erlbaum.

Anderson, J. R. (2020). *Cognitive psychology and its implications* (9th ed.). Worth Publishers.

Anderson, J. R., Betts, S., Bothell, D., Hope, R., & Lebiere, C. (2019). Learning rapid and precise skills. *Psychological Review*, *126*(5), 727–763.

Anderson, J. R., Betts, S., Bothell, D., & Lebiere, C. (2021). Discovering skill. *Cognitive Psychology*, *129*, 101417.

Anderson, J. R., & Fincham, J. M. (2014). Extending problem-solving procedures through reflection. *Cognitive Psychology*, *74*, 1–34.

Anderson, J. R., & Lebiere, C. (1998). *The atomic components of thought*. Erlbaum.

Andreu, C. I., Palacios, I., Moënne-Loccoz, C., López, V., Franken, I. H., Cosmelli, D., & Slagter, H. A. (2019). Enhanced response inhibition and reduced midfrontal theta activity in experienced Vipassana meditators. *Scientific Reports*, *9*(1), 12870.

Arango-Muñoz, S. (2011). Two levels of metacognition. *Philosophia*, *39*(1), 71–82.

Awh, E., Belopolsky, A. V., & Theeuwes, J. (2012). Top-down versus bottom-up attentional control: A failed theoretical dichotomy. *Trends in Cognitive Sciences*, *16*(8), 437–443.

Baird, B., Mrazek, M. D., Phillips, D. T., & Schooler, J. W. (2014). Domain-specific enhancement of metacognitive ability following meditation training. *Journal of Experimental Psychology: General*, *143*(5), 1972–1979.

Baker, L., & Brown, A. L. (1984). Metacognitive skills and reading. In P. D. Pearson (Ed.), *Handbook of reading research* (pp. 353–394). Longman.

Baumeister, R. F. (1984). Choking under pressure: self-consciousness and paradoxical effects of incentives on skillful performance. *Journal of personality and social psychology,* 46(3), 610.

Beilock, S. L., Afremow, J. A., Rabe, A. L., & Carr, T. H. (2001). "Don't miss!" The debilitating effects of suppressive imagery on golf putting performance. *Journal of Sport & Exercise Psychology*, *23*(3), 200–221.

Beilock, S. L., & Carr, T. H. (2001). On the fragility of skilled performance: What governs choking under pressure? *Journal of Experimental Psychology: General*, *130*(4), 701–725.

Beilock, S. L., & Carr, T. H. (2004). From novice to expert performance: Memory, attention and the control of complex sensori-motor skills. In A. M. Williams & N. J. Hodges (Eds.), *Skill acquisition in sport: Research, theory and practice* (pp. 309–328). Routledge.

Beilock, S. L., & Carr, T. H. (2005). When high-powered people fail: Working memory and "choking under pressure" in math. *Psychological Science*, *16*(2), 101–105.

Bergamaschi Ganapini, M., Campbell, M., Fabiano, F., Horesh, L., Lenchner, J., Loreggia, A., ... & Venable, K. B. (2025). Fast, slow, and metacognitive thinking in AI. *npj Artificial Intelligence*, *1*(1), 27.

Bolzenkötter, T., Bürkner, P. C., Zetsche, U., & Schulze, L. (2024). Assessing the immediate effects of detached mindfulness on repetitive negative thinking and affect in daily life: A randomized controlled trial. *Mindfulness*, *15*(5), 1136-1148.

Brown, K. W., & Ryan, R. M. (2003). The benefits of being present: Mindfulness and its role in psychological well-being. *Journal of Personality and Social Psychology*, *84*(4), 822–848.

Cardaciotto, L., Herbert, J. D., Forman, E. M., Moitra, E., & Farrow, V. (2008). The assessment of present-moment awareness and acceptance: The Philadelphia Mindfulness Scale. *Assessment*, *15*(2), 204–223.

Carpenter, J., Sherman, M. T., Kievit, R. A., Seth, A. K., Lau, H., & Fleming, S. M. (2019). Domain-general enhancements of metacognitive ability through adaptive training. *Journal of Experimental Psychology: General*, *148*(1), 51–64.

Carruthers, P., & Williams, D. M. (2022). Model-free metacognition. *Cognition*, *225*, 105117.

Chambers, R., Lo, B., & Allen, N. B. (2008). The impact of intensive mindfulness training on attentional control, cognitive style, and affect. *Cognitive Therapy and Research*, *32*(3), 303–322.

Charles, L., Chardin, C., & Haggard, P. (2020). Evidence for metacognitive bias in perception of voluntary action. *Cognition*, *203*, 104335.

Charlton, S., & Starkey, N. (2013). Driving on familiar roads: Automaticity and inattention blindness. *Transportation Research Part F: Traffic Psychology and Behaviour*, *19*, 121–133.

Choi, J., Cho, H., Choi, J. S., Choi, I. Y., Chun, J. W., & Kim, D. J. (2021). The neural basis underlying impaired attentional control in problematic smartphone users. *Translational Psychiatry*, *11*(1), 129.

Christensen, W., Sutton, J., & McIlwain, D. J. (2016). Cognition in skilled action: Meshed control and the varieties of skill experience. *Mind & Language*, *31*(1), 37–66.

Chua, L.-K., Jimenez-Diaz, J., Lewthwaite, R., Kim, T., & Wulf, G. (2021). Superiority of external attentional focus for motor performance and learning: Systematic reviews and meta-analyses. *Psychological Bulletin*, *147*(6), 618–645.

Clark, A. (2013). *Mindware: An Introduction to the Philosophy of Cognitive Science.* Oxford University Press.

Clark, A. (2015). *Surfing uncertainty: Prediction, action, and the embodied mind*. Oxford University Press.

Conway-Smith, B. (2025). Metacognition as a domain of skill. In *Proceedings of the Annual Conference of the Cognitive Science Society* (Vol. 47, pp. 2646–2653).

Conway-Smith, B. (2026). *Computational models of metacognitive monitoring and control* [Doctoral dissertation]. GitHub. https://github.com/BrendanCS/PhD_thesis_models

Conway-Smith, B., & West, R. (2024). Toward autonomy: Metacognitive learning for enhanced AI performance. In *AAAI Spring Symposium on Human-Like Learning in AI* (pp. 545–546).

Conway-Smith, B., & West, R. (2025). Metacognition in HCI: Designing systems for planning and flexibility. In R. A. Sottilare & J. Schwarz (Eds.), *Adaptive Instructional Systems* (Vol. 15812, pp. 176–187). Springer.

Conway-Smith, B., & West, R. L. (2022). System-1 and System-2 realized within the Common Model of Cognition. In *AAAI Fall Symposium: Thinking Fast & Slow and Other Cognitive Theories in AI*.

Conway-Smith, B., & West, R. L. (2023). Metacognitive threshold: A computational account. In *Proceedings of ICCM 2023: 21st International Conference on Cognitive Modeling* (pp. 70–75).

Conway-Smith, B., & West, R. L. (2024). The computational mechanisms of detached mindfulness. In *Proceedings of ICCM 2024: 22nd International Conference on Cognitive Modeling* (pp. 43–49).

Conway-Smith, B., & West, R. L. (2025). Metacognitive mechanisms of the attentional training technique. In *Proceedings of ICCM 2025: 23rd International Conference on Cognitive Modeling* (pp. 14–20).

Conway-Smith, B., West, R. L., & Mylopoulos, M. (2023). Metacognitive skill: How it is acquired. In *Proceedings of the Annual Conference of the Cognitive Science Society* (Vol. 45, pp. 2646–2653).

Cooney, R. E., Joorman, J., Eugene, F., Dennis, E. L., & Gotlib, I. H. (2010). Neural correlates of rumination in depression. *Cognitive, Affective, & Behavioral Neuroscience*, *10*(4), 470–478.

Cross, D. R., & Paris, S. G. (1988). Developmental and instructional analyses of children's metacognition. *Journal of Educational Psychology*, *80*(2), 131–142.

Cohen, N. J., & Squire, L. R. (1980). Preserved learning and retention of pattern-analyzing skill in amnesia: Dissociation of knowing how and knowing that. *Science*, 210(4466), 207–210.

Corbetta, M., & Shulman, G. L. (2002). Control of goal-directed and stimulus-driven attention in the brain. *Nature Reviews Neuroscience, 3*(3), 201–215.

Davidson, J. E., & Sternberg, R. J. (1998). Smart problem solving: How metacognition helps. In D. J. Hacker, J. Dunlosky, & A. C. Graesser (Eds.), *Metacognition in educational theory and practice* (pp. 47–68). Routledge.

Dayan, P. (2009). Goal-directed control and its antipodes. *Neural Networks*, *22*(8), 1078–1089.

de Boer, H., Donker, A. S., Kostons, D. D. N. M., & van der Werf, G. P. C. (2018). Long-term effects of metacognitive strategy instruction on student academic performance: A meta-analysis. *Educational Research Review*, *24*, 98–115.

de Groot, A. D. (1978). *Thought and choice in chess* (2nd ed.). Mouton Publishers.

Dennett, D. C. (1978). *Brainstorms: Philosophical essays on mind and psychology*. MIT Press.

Dobson, K. S. (2013). The science of CBT: Toward a metacognitive model of change? *Behavior Therapy*, *44*(2), 224–237.

Douskos, C. (2019). The spontaneousness of skill and the impulsivity of habit. *Synthese*, *196*(11), 4513–4535.

Dreyfus, H. L., & Dreyfus, S. E. (1986). *Mind over machine: The power of human intuition and expertise in the era of the computer*. Free Press.

Ducrocq, E., Wilson, M. R., Vine, S. J., & Derakshan, N. (2016). Training attentional control improves cognitive and motor task performance. *Journal of Sport and Exercise Psychology*, *38*(5), 521–533.

Dunlosky, J., & Rawson, K. A. (2019). How cognitive psychology can inform evidence-based education reform. In J. Dunlosky & K. A. Rawson (Eds.), *The Cambridge handbook of cognition and education* (pp. 1–22). Cambridge University Press.

Dunlosky, J., Rawson, K. A., Marsh, E. J., Nathan, M. J., & Willingham, D. T. (2013). Improving students' learning with effective learning techniques: Promising directions from cognitive and educational psychology. *Psychological Science in the Public Interest, 14*(1), 4–58.

Eberth, J., Sedlmeier, P., & Schäfer, T. (2019). PROMISE: A model of insight and equanimity as the key effects of mindfulness meditation. *Frontiers in Psychology*, *10*, 272.

Elliott, L. J., Lum, H. C., Aqlan, F., Zhao, R., & Lasher, C. D. (2019, June). A study of metacognitive problem solving in undergraduate engineering students. In *International Conference on Applied Human Factors and Ergonomics* (pp. 95–102). Springer.

Ellis, R. (1994). *The study of second language acquisition*. Oxford University Press.

Fazio, L. K., Brashier, N. M., Payne, B. K., & Marsh, E. J. (2015). Knowledge does not protect against illusory truth. *Journal of experimental psychology: general*, *144*(5), 993.

Fisher, P. L. (2021). Metacognitive therapy. In S. G. Hofmann & G. J. G. Asmundson (Eds.), *Handbook of cognitive behavioral therapy: Overview and approaches* (Vol. 1, pp. 617–636). American Psychological Association.

Fitts, P. M. (1964). Perceptual-motor skill learning. In A. W. Melton (Ed.), *Categories of human learning* (pp. 243–285). Academic Press.

Fitts, P. M., & Posner, M. I. (1967). *Human performance*. Brooks/Cole.

Flavell, J. H. (1979). Metacognition and cognitive monitoring: A new area of cognitive–developmental inquiry. *American Psychologist*, *34*(10), 906–911.

Fletcher, L., & Carruthers, P. (2012). Metacognition and reasoning. *Philosophical Transactions of the Royal Society B: Biological Sciences, 367*(1594), 1366–1378.

Fleming, S. M., Dolan, R. J., & Frith, C. D. (2012). Metacognition: Computation, biology and function. *Philosophical Transactions of the Royal Society B: Biological Sciences*, *367*(1594).

Fleming, S. M., & Lau, H. C. (2014). How to measure metacognition. *Frontiers in Human Neuroscience*, *8*, 443.

Fodor, J. A. (1975). *The language of thought.* Harvard University Press.

Ford, B. Q., Gross, J. J., & Gruber, J. (2019). Broadening our field of view: The role of emotion polyregulation. *Emotion Review*, *11*(3), 197-208.

Ford, P. R., Hodges, N. J., & Williams, A. M. (2005). Online attentional-focus manipulations in a soccer-dribbling task: Implications for the proceduralization of motor skills. *Journal of Motor Behavior*, *37*(5), 386–394.

Fox, K., Dixon, M., Nijeboer, S., Girn, M., Floman, J., Lifshitz, M., & Christoff, K. (2016). Functional neuroanatomy of meditation: A review and meta-analysis of 78 functional neuroimaging investigations. *Neuroscience & Biobehavioral Reviews*, *65*, 208–228.

Fridland, E. (2017). Skill and motor control: Intelligence all the way down. *Philosophical Studies,* 174, 1539–1560.

Fridland, E. (2019). Longer, smaller, faster, stronger: On skills and intelligence. *Philosophical Psychology*, *32*(3), 369–386.

Garofalo, J., & Lester, F. K. (1985). Metacognition, cognitive monitoring, and mathematical performance. *Journal for Research in Mathematics Education*, *16*(3), 163–176.

Girash, J. (2014). Metacognition and instruction. In V. A. Benassi, C. E. Overson, & C. M. Hakala (Eds.), *Applying science of learning in education: Infusing psychological science into the curriculum* (pp. 152–168). Society for the Teaching of Psychology.

Gross, J. J. (2014). Emotion regulation: Conceptual and empirical foundations. In J. J. Gross (Ed.), *Handbook of emotion regulation* (2nd ed., pp. 3–20). Guilford Press.

Gross, J. J. (2015). Emotion regulation: Current status and future prospects. *Psychological Inquiry*, *26*(1), 1-26.

Grossman, P. (2010). Mindfulness for psychologists: Paying kind attention to the perceptible. *Mindfulness,* 1, 87–97.

Grossman, P., Niemann, L., Schmidt, S., & Walach, H. (2004). Mindfulness-based stress reduction and health benefits: A meta-analysis. Journal of Psychosomatic Research, 57(1), 35–43.

Güner, P., & Erbay, H. N. (2021). Metacognitive skills and problem-solving. *International Journal of Research in Education and Science, 7*(3), 715–734.

Gutierrez de Blume, A. P. (2022). Calibrating calibration: A meta-analysis of learning strategy instruction interventions to improve metacognitive monitoring accuracy. *Journal of Educational Psychology, 114*(4), 681–700.

Hagen, R., Hjemdal, O., Solem, S., Kennair, L. E. O., Nordahl, H. M., Fisher, P., & Wells, A. (2017). Metacognitive therapy for depression in adults: A waiting list randomized controlled trial with six months follow-up. *Frontiers in Psychology*, *8*, 31.

Hammersmark, A. T., Hjemdal, O., Hannisdal, M., Lending, H., Reme, S., Hodne, K., Osnes, K., Gjengedal, R., & Johnson, S. U. (2024). Metacognitive therapy for generalized anxiety disorders in group: A case study. *Journal of Clinical Psychology, 80*(4), 884–899.

Hart, J. T. (1965). Memory and the feeling-of-knowing experience. *Journal of Educational Psychology*, *56*(4), 208–216.

Hameed, H. A., & Cheruvalath, R. (2021). Metacognitive skills inventory (MSI): Development and validation. *International Journal of Testing*, *21*(2), 144–167.

Holas, P., & Jankowski, T. (2013). A cognitive perspective on mindfulness. *International Journal of Psychology,* 48(3), 232-243.

Jäger, C., & Bänninger-Huber, E. (2015). Looking into meta-emotions. Phenomenology and the Cognitive Sciences, 14(1), 171–188.

Jahn, N., Sinke, C., Kayali, Ö., Krug, S., Leichter, E., Peschel, S., & Heitland, I. (2023). Neural correlates of the attention training technique as used in metacognitive therapy: A randomized sham-controlled fMRI study in healthy volunteers. *Frontiers in Psychology*, 14, 1140688.

Jankowski, T., & Holas, P. (2014). Metacognitive model of mindfulness. *Consciousness and cognition,* 28, 64-80.

Johnson, S. G. B., Karimi, A.-H., Bengio, Y., Chater, N., Gerstenberg, T., Larson, K., Levine, S., Mitchell, M., Rahwan, I., Schölkopf, B., & Grossmann, I. (2026). Imagining and building wise machines: The centrality of AI metacognition. *Trends in Cognitive Sciences.*

Juvina, I., Larue, O., & Hough, A. (2018). Modeling valuation and core affect in a cognitive architecture: The impact of valence and arousal on memory and decision-making. *Cognitive Systems Research*, *48*, 4-24.

Kahneman, D., & Frederick, S. (2004). Attribute substitution in intuitive judgment. In *Models of a man: Essays in memory of Herbert A. Simon* (pp. 49–81). Psychology Press.

Kakumanu, R. J., Nair, A., Venugopal, R., Sasidharan, A., Ghosh, P., John, J. P., Mehrotra, S., Panth, R., & Kutty, B. M. (2018). Dissociating meditation proficiency and experience dependent EEG changes during traditional Vipassana meditation practice. *Biological Psychology*, 135, 65–75.

Keith, N., & Frese, M. (2005). Self-regulation in error management training: Emotion control and metacognition as mediators of performance effects. *Journal of Applied Psychology* (4), 677–691.

Kingdom, F. A. A., & Prins, N. (2016). *Psychophysics: A practical introduction* (2nd ed.). Academic Press.

Kim, J. W., & Ritter, F. E. (2015). Learning, forgetting, and relearning for keystroke- and mouse-driven tasks: Relearning is important. *Human–Computer Interaction*, *30*(2), 155–189.

Kahneman, D., & Klein, G. (2009). Conditions for intuitive expertise: a failure to disagree. *American psychologist,* 64(6), 515.

Knowles, M. M., Foden, P., El-Deredy, W., & Wells, A. (2016). A systematic review of efficacy of the attention training technique in clinical and nonclinical samples. *Journal of Clinical Psychology*, *72*(10), 999–1015.

Knowles, M. & Wells, A. (2018). Single dose of the attention training technique increases resting alpha and beta oscillations in frontoparietal brain networks. *Frontiers in Psychology*, *9*, 272.

Koriat, A. (2000). The feeling of knowing: Some metatheoretical implications for consciousness and control. *Consciousness and Cognition*, *9*(2), 149–171.

Koriat, A., & Levy-Sadot, R. (1999). Information-based and experience-based monitoring of one's own knowledge. In S. Chaiken & Y. Trope (Eds.), *Dual-process theories in social psychology* (pp. 483–502). Guilford Press.

Kotseruba, I., & Tsotsos, J. K. (2020). 40 years of cognitive architectures: Core cognitive abilities and practical applications. *Artificial Intelligence Review*, 53, 17–94

Kralik, J. D., Lee, J. H., Rosenbloom, P. S., Jackson Jr, P. C., Epstein, S. L., Romero, O. J., ... & McGreggor, K. (2018). Metacognition for a common model of cognition. *Procedia Computer Science,* 145, 730-739.

Krueger, P. M., Lieder, F., & Griffiths, T. (2017). Enhancing metacognitive reinforcement learning using reward structures and feedback. In *Proceedings of the Annual Conference of the Cognitive Science Society*.

Kuhn, D., & Dean, D. (2004). Metacognition: A bridge between cognitive psychology and educational practice. *Theory Into Practice*, *43*(4), 268–273.

Kuhn, T. S. (1962). *The structure of scientific revolutions*. University of Chicago Press.

Kuhn, D., & Pearsall, S. (2000). Developmental origins of scientific thinking. *Journal of Cognition and Development,* 1(1), 113-129.

Laird, J. E. (2012). *The Soar cognitive architecture*. MIT Press.

Laird, J. E., Lebiere, C., & Rosenbloom, P. S. (2017). A standard model of the mind: Toward a common computational framework across artificial intelligence, cognitive science, neuroscience, and robotics. *AI Magazine*, *38*(4), 13–26.

Landis, C. (1954). Determinants of the critical flicker-fusion threshold. *Physiological Reviews*, *34*(2), 259–286.

Lange, V. (2025a). A reductive account of mindfulness as metacognitive control. *Review of Philosophy and Psychology*, *16*, 1–25.

Lange, V. (2025b). Decentering and attention. *Philosophical Psychology*, 38(4), 1530-1557.

Leder, G., & Zawidzki, T. (2023). The skill of mental health: Towards a new theory of mental health and disorder. *Philosophy and the Mind Sciences*, *4*, 1–29.

Lewis, B., & Linder, D. (1997). Thinking about choking? Attentional processes and paradoxical performance. *Personality and Social Psychology Bulletin*, *23*(9), 937–944.

Lieder, F., Shenhav, A., Musslick, S., & Griffiths, T. L. (2018). Rational metareasoning and the plasticity of cognitive control. *PLOS Computational Biology*, *14*(4), e1006043.

Logan, G. D. (1988). Toward an instance theory of automatization. *Psychological Review,* 95(4), 492.

Lutz, A., Slagter, H. A., Dunne, J. D., & Davidson, R. J. (2008). Attention regulation and monitoring in meditation. *Trends in Cognitive Sciences*, *12*(4), 163–169.

Marr, D. (1982). *Vision: A computational investigation into the human representation and processing of visual information*. W. H. Freeman.

McCabe, J. (2011). Metacognitive awareness of learning strategies in undergraduates. *Memory & cognition, 3*9(3), 462-476.

McCormick, C. B. (2003). Metacognition and learning. *Handbook of psychology*, 79-102.

MacLean, K. A., Ferrer, E., Aichele, S. R., Bridwell, D. A., Zanesco, A. P., Jacobs, T. L., & Saron, C. D. (2010). Intensive meditation training improves perceptual discrimination and sustained attention. *Psychological Science*, *21*(6), 829–839.

Masters, R. S. W. (1992). Knowledge, nerves and know-how: The role of explicit versus implicit knowledge in the breakdown of a complex motor skill under pressure. *British Journal of Psychology*, *83*(3), 343–358.

Meher, V., Baral, R., & Bhuyan, S. (2021). A meta-analysis on the effectiveness of metacognitive strategies and interventions in teaching and learning process. *i-manager's Journal on Educational Psychology*, *14*(4), 47–58.

Moore, R. G. (1996). It's the thought that counts: The role of intentions and meta-awareness in cognitive therapy. *Journal of Cognitive Psychotherapy*, 10(4), 255–269.

Mylopoulos, M., & Pacherie, E. (2020). Self-control as hybrid skill. In R. Shaffer & J. W. Schooler (Eds.), *Surrounding self-control* (pp. 81–100). Oxford University Press.

Mylopoulos, M., & Pacherie, E. (2021). Skilled action control. *Review of Philosophy and Psychology*, *12*(4), 797–820.

Nassif, Y., & Wells, A. (2014). Attention training reduces intrusive thoughts cued by a narrative of stressful life events: A controlled study. *Journal of Clinical Psychology,* 70(6), 510–517.

Nelson, T. O., & Narens, L. (1990). Metamemory: A theoretical framework and new findings. In G. H. Bower (Ed.), *The psychology of learning and motivation* (Vol. 26). Academic Press.

Newell, A. (1973). You can't play 20 questions with nature and win: Projective comments on the papers of this symposium. In W. G. Chase (Ed.), *Visual information processing: Proceedings of the eighth annual Carnegie Symposium on Cognition* (pp. 283–308). Academic Press.

Newell, A. (1990). *Unified theories of cognition*. Harvard University Press.

Newell, A., & Rosenbloom, P. S. (1981). Mechanisms of skill acquisition and the law of practice. In J. R. Anderson (Ed.), *Cognitive skills and their acquisition* (pp. 1–55). Erlbaum.

Newell, A., & Simon, H. A. (1976). Computer science as empirical inquiry: Symbols and search. *Communications of the ACM, 19*(3), 113–126.

Normann, N., & Morina, N. (2018). The efficacy of metacognitive therapy: A systematic review and meta-analysis. *Frontiers in Psychology*, *9*, 2211.

Oxford, R. L. (2011). *Strategies for learning a second or foreign language.* Language teaching, 44(2), 167-180.

Pacherie, E., & Mylopoulos, M. (2021). Beyond automaticity: The psychological complexity of skill. *Topoi,* 40(3), 649-662.

Pauen, M., & Haynes, J. (2021). Measuring the mental. *Consciousness and Cognition,* 90, 103106.

Pavese, C. (2019). The psychological reality of practical representation. *Philosophical Psychology*, *32*(3), 387–411.

Pearman, A., Lustig, E., Hughes, M. L., & Hertzog, C. (2020). Initial evidence for the efficacy of an everyday memory and metacognitive intervention. *Innovation in Aging*, *4*(3), igaa020.

Pedone, R., Semerari, A., Riccardi, I., Procacci, M., & Carcione, A. (2017). Development of a self-report measure of metacognition: The Metacognition Self-Assessment Scale (MSAS). *Clinical Neuropsychiatry*, *14*(3), 185–194.

Posner, M. I., Rothbart, M. K., & Tang, Y.-Y. (2015). Enhancing attention through training. In *Cognitive enhancement* (pp. 213–231). Elsevier.

Predatu, R., David, D. O., & Maffei, A. (2020). Beliefs about emotions, negative meta-emotions, and perceived emotional control during an emotionally salient situation in individuals with emotional disorders. Cog*nitive Therapy and Research,* 44(2), 287–299.

Prins, N. (2016). *Psychophysics: A practical introduction*. Academic Press.

Proust, J. (2013). *The philosophy of metacognition: Mental agency and self-awareness*. Oxford University Press.

Proust, J. (2019). From comparative studies to interdisciplinary research on metacognition. *Animal Behavior and Cognition*, *6*(4), 285–302.

Pylyshyn, Z. W. (1984). *Computation and cognition: Toward a foundation for cognitive science.* MIT Press.

Racsmány, M., Szőllősi, Á., & Bencze, D. (2018). Retrieval practice makes procedure from remembering: An automatization account of the testing effect. *Journal of Experimental Psychology: Learning, Memory, and Cognition*, *44*(1), 157.

Ramamurthy, M., & Blaser, E. (2017). New rules for visual selection: Isolating procedural attention. *Journal of Vision*, *17*(6), 15.

Richards, J. M., & Gross, J. J. (2000). Emotion regulation and memory: The cognitive costs of keeping one's cool. *Journal of Personality and Social Psychology*, *79*(3), 410–424.

Rigby, C. S., Schultz, P. P., & Ryan, R. M. (2014). Mindfulness, interest-taking, and self-regulation. In A. Ie, C. T. Ngnoumen, & E. J. Langer (Eds.), *The Wiley Blackwell handbook of mindfulness* (pp. 775–798). Wiley Blackwell.

Ritter, F. E., Reifers, A., Klein, A., & Schoelles, M. (2006). Lessons from defining theories of stress. In W. Gray (Ed.) *Integrated Models of Cognitive Systems.* New York: Oxford University Press.

Rochat, L., Manolov, R., & Billieux, J. (2018). Efficacy of metacognitive therapy in improving mental health: A meta-analysis of single-case studies. *Journal of Clinical Psychology*, *74*(12).

Rouder, J. N., & Morey, R. D. (2009). The nature of psychological thresholds. *Psychological Review*, *116*(3), 655–660.

Ryle, G. (1949). *The concept of mind*. Hutchinson.

Salovich, N. A., Remington, R. W., & Jiang, Y. V. (2018). Acquisition of habitual visual attention and transfer to related tasks. *Psychonomic Bulletin & Review*, *25*(1), 272–278.

Schaeffner, S., Chevalier, N., Kubota, M., & Karbach, J. (2021). Metacognitive training. In *Cognitive training: An overview of features and applications* (pp. 247–274). Springer.

Schneider, W., & Shiffrin, R. M. (1977). Controlled and automatic human information processing: I. Detection, search, and attention. *Psychological Review*, 84(1), 1–66.

Schraw, G., & Dennison, R. S. (1994). Metacognitive Awareness Inventory. *Contemporary Educational Psychology*, *19*(4), 460–475.

Schraw, G., & Moshman, D. (1995). Metacognitive theories. *Educational Psychology Review*, *7*(4), 351–371.

Schraw, G., Crippen, K. J., & Hartley, K. (2006). Promoting self-regulation in science education: Metacognition as part of a broader perspective on learning. *Research in Science Education*, *36*(1), 111–139.

Schuster, C., Stebner, F., Leutner, D., & Wirth, J. (2020). Transfer of metacognitive skills in self-regulated learning: An experimental training study. *Metacognition and Learning*, *15*(3).

Shea, N. (2024a). *Concepts at the interface*. Oxford University Press.

Shea, N. (2024b). Metacognition of inferential transitions. *Journal of Philosophy, 121*(11).

Shea, N., Boldt, A., Bang, D., Yeung, N., Heyes, C., & Frith, C. D. (2014). Supra-personal cognitive control and metacognition. *Trends in Cognitive Sciences*, *18*(4), 186–193.

Shepherd, J. (2021). *The shape of agency: Control, action, skill, knowledge*. Oxford University Press.

Sherman, M. T., Seth, A. K., & Kanai, R. (2018). Quantifying metacognitive thresholds using signal-detection theory. *bioRxiv.*

Shin, J. C., Chang, S., & Cho, Y. S. (2015). Adjustment to subtle time constraints and power law learning in rapid serial visual presentation. *Frontiers in Psychology,* 6, 1748.

Singley, M. K., & Anderson, J. R. (1989). *The transfer of cognitive skill*. Harvard University Press.

Slagter, H. A., Davidson, R. J., & Lutz, A. (2011). Mental training as a tool in the neuroscientific study of brain and cognitive plasticity. *Frontiers in Human Neuroscience*, *5*, 17.

Squire, L. R. (1992). Declarative and nondeclarative memory: Multiple brain systems supporting learning and memory. *Journal of Cognitive Neuroscience*, *4*(3), 232–243.

Squire, L. R., & Zola, S. M. (1996). Structure and function of declarative and nondeclarative memory systems. *Proceedings of the National Academy of Sciences*, *93*(24), 13515–13522.

Stanley, J., & Williamson, T. (2001). Knowing how. *The Journal of Philosophy*, *98*(8), 411–444.

Stanton, J. D., Sebesta, A. J., & Dunlosky, J. (2021). Fostering metacognition to support student learning and performance. *CBE—Life Sciences Education*, 20(2).

Stewart, T. C., Choo, X., & Eliasmith, C. (2010). Dynamic behaviour of a spiking model of action selection in the basal ganglia. In *Proceedings of the 10th International Conference on Cognitive Modeling,* 235–240.

Stocco, A. (2018). A biologically plausible action selection system for cognitive architectures: Implications of basal ganglia anatomy for learning and decision-making models. *Cognitive Science*, *42*(2), 564–607.

Stocco, A., Sibert, C., Steine-Hanson, Z., Koh, N., Laird, J. E., Lebiere, C. J., & Rosenbloom, P. (2021). Analysis of the human connectome data supports the notion of a "Common Model of Cognition" for human and human-like intelligence across domains. *NeuroImage*, *235*, 118035.

Sun, R. (2006). The CLARION cognitive architecture: Extending cognitive modeling to social simulation. *Cognition and multi-agent interaction,* 79-99.

Sun, R., Merrill, E., & Peterson, T. (2001). From implicit skills to explicit knowledge: A bottom-up model of skill learning. *Cognitive Science*, *25*(2), 203–244.

Taatgen, N. (2013). The nature and transfer of cognitive skills. *Psychological Review*, *120*(3), 439–471.

Taatgen, N. A., & Lee, F. J. (2003). Production compilation: A simple mechanism to model complex skill acquisition. *Human Factors*, *45*(1), 61–76.

Tang, Y.-Y. (2017). *The neuroscience of mindfulness meditation: How the body and mind work together to change our behaviour.* Springer International Publishing.

Tang, Y.-Y., Hölzel, B. K., & Posner, M. I. (2015). The neuroscience of mindfulness meditation. *Nature Reviews Neuroscience, 16*(4), 213–225.

Tenison, C., & Anderson, J. R. (2016). Modeling the distinct phases of skill acquisition. *Journal of Experimental Psychology: Learning, Memory, and Cognition*, *42*(5), 749–767.

Theeuwes, J. (2010). Top–down and bottom–up control of visual selection. *Acta Psychologica, 135*(2), 77–99.

Thompson, V. A. (2009). Dual-process theories: A metacognitive perspective. In J. Evans & K. Frankish (Eds.), *In two minds: Dual processes and beyond.* Oxford University Press.

Tuncer, M., & Kaysi, F. (2013). The development of the metacognitive thinking skills scale. *Development*, *2*(1), 1–9.

Ukov, T., & Tsochev, G. (2025). Reviewing a model of metacognition for application in cognitive architecture design. *Systems*, 13(3), 177.

Van Dam, N. T., Van Vugt, M. K., Vago, D. R., Schmalzl, L., Saron, C. D., Olendzki, A., & Meyer, D. E. (2018). Mind the hype: A critical evaluation and prescriptive agenda for research on mindfulness and meditation. *Perspectives on Psychological Science*, *13*(1), 36–61.

Van der Stel, M., & Veenman, M. V. (2010). Development of metacognitive skillfulness: A longitudinal study. *Learning and Individual Differences*, *20*(3), 220–224.

Veenman, M. V. (2015). Metacognition. In *Handbook of individual differences in reading* (pp. 95–114). Routledge.

Veenman, M. V. J. (2017). Assessing metacognitive deficiencies and effectively instructing metacognitive skills. *Teachers College Record,* 119(13), 1–20.

Veenman, M., & Elshout, J. J. (1999). Changes in the relation between cognitive and metacognitive skills during the acquisition of expertise. *European Journal of Psychology of Education*, *14*(4), 509–523.

Veenman, M. V., & Spaans, M. A. (2005). Relation between intellectual and metacognitive skills: Age and task differences. *Learning and Individual Differences*, *15*(2), 159–176.

Veenman, M. V., Van Hout-Wolters, B. H., & Afflerbach, P. (2006). Metacognition and learning: Conceptual and methodological considerations. *Metacognition and Learning*, *1*(1), 3–14.

Vohs, K. D., & Baumeister, R. F. (2016). *Handbook of self-regulation: Research, theory, and applications*. Guilford Publications.

Wadlinger, H. A., & Isaacowitz, D. M. (2011). Fixing our focus: Training attention to regulate emotion. *Personality and Social Psychology Review*, *15*(1), 75–102.

Walker, P. B., Haase, J. J., Mehalick, M. L., Steele, C. T., Russell, D. W., & Davidson, I. N. (2025). Harnessing metacognition for safe and responsible AI. *Technologies*, *13*(3), 107.

Wang, J., Marchant, D., Morris, T., & Gibbs, P. (2004). Self-consciousness and trait anxiety as predictors of choking in sport. *Journal of Science and Medicine in Sport*, *7*(2), 174–185.

Wei, H., Shakarian, P., Lebiere, C., Draper, B., Krishnaswamy, N., & Nirenburg, S. (2024). *Metacognitive AI: Framework and the Case for a Neurosymbolic Approach.* arXiv.

Wells, A. (1990). Panic disorder in association with relaxation induced anxiety: An attentional training approach to treatment. *Behavior Therapy*, *21*(3), 273–280.

Wells, A. (1995). Meta-cognition and worry: A cognitive model of generalized anxiety disorder. *Behavioural and Cognitive Psychotherapy,* 23(3), 301–320.

Wells, A. (2000). *Emotional disorders and metacognition: Innovative cognitive therapy.* Wiley.

Wells, A. (2005). Detached mindfulness in cognitive therapy: A metacognitive analysis and ten techniques. *Journal of Rational-Emotive & Cognitive-Behavior Therapy,* 23(4), 337–355.

Wells, A. (2009). *Metacognitive therapy for anxiety and depression*. Guilford Press.

Wells, A. (2019). Breaking the cybernetic code: Understanding and treating the human metacognitive control system to enhance mental health. *Frontiers in Psychology*, *10*, 2626.

Wells, A., & Matthews, G. (1994). Self-consciousness and cognitive failures as predictors of coping in stressful episodes. *Cognition & Emotion*, *8*(3), 279–295.

Wells, A., & Matthews, G. (1996). Modelling cognition in emotional disorder: The S-REF model. *Behaviour Research and Therapy*, *34*(11–12), 881–888.

West, R. L., & Conway-Smith, B. (2019). Put feeling into cognitive models: A computational theory of feeling. In *Proceedings of ICCM 2019: 17th International Conference on Cognitive Modeling (pp. 295–300).*

Wongpakaran, N., Wongpakaran, T., Wedding, D., Mirnics, Z., & Kövi, Z. (2021, September). Role of equanimity on the mediation model of neuroticism, perceived stress and depressive symptoms. In *Healthcare (*Vol. 9, No. 10, p. 1300). MDPI.

World Health Organization. (2022). *World mental health report: Transforming mental health for all*. World Health Organization.

Wu, S., Oltramari, A., & Ritter, F. E. (2025). VSM-ACTR 2: a human-like decision making model with metacognition for manufacturing solutions. *Computational and Mathematical Organization Theory*, 31(4), 259-276.

Zawidzki, T. (2019). Metacognitive skill and the therapeutic regulation of emotion. *Philosophical Topics*, *47*(2), 27–52.

Zimmerman, B. J., & Schunk, D. H. (2011). Self-regulated performance: An introduction and overview. In D. H. Schunk & B. J. Zimmerman (Eds.), *Handbook of self-regulation of learning and performance* (pp. 1–12). Routledge.